\documentclass[11pt,a4paper]{scrartcl}

\usepackage{subfig}

\usepackage{CLICdp}

\usepackage{CLICdp_definitions}
\usepackage[bottom]{footmisc}
\usepackage{graphicx}

\usepackage{float}

\newcommand{\subfigimg}[3][,]{%
  \setbox1=\hbox{\includegraphics[#1]{#3}}% Store image in box
  \leavevmode\rlap{\usebox1}% Print image
  \rlap{\hspace*{-16pt}\raisebox{\dimexpr\ht1-1\baselineskip}{#2}}% Print label
  \phantom{\usebox1}% Insert appropriate spcing
}

\title{Tracking performance and simulation of capacitively coupled pixel detectors for the CLIC vertex detector}

\clicdppub{2018}{006}  % journal articles
\date{\today}
\addauthor{N.~Alipour~Tehrani}{\institute{1}}
\addauthor{M.~Benoit}{\institute{1}\institute{2}}
\addauthor{M.~Buckland\thanks{Corresponding author \newline Email address: buckland@liverpool.ac.uk}}{\institute{3}\institute{1}}
\addauthor{D.~Dannheim}{\institute{1}}
\addauthor{A.~Fiergolski}{\institute{1}}
\addauthor{S.~Green}{\institute{4}}
\addauthor{D.~Hynds\thanks{Now at NIKHEF, The Netherlands}}{\institute{1}}
\addauthor{I.~Kremastiotis}{\institute{1}\institute{5}}
\addauthor{S.~Kulis}{\institute{1}}
\addauthor{M.~Munker}{\institute{1}}
\addauthor{A.~N{\"u}rnberg}{\institute{1}}
\addauthor{I.~Peric}{\institute{5}}
\addauthor{M.~Petric}{\institute{1}}
\addauthor{E.~Sicking}{\institute{1}}
\addauthor{M.~Vicente}{\institute{1}\institute{2}}

\addinstitute{1}{CERN, Switzerland}
\addinstitute{2}{University of Geneva, Switzerland}
\addinstitute{3}{University of Liverpool, United Kingdom}
\addinstitute{4}{University of Cambridge, United Kingdom}
\addinstitute{5}{Karlsruhe Institute of Technology, Germany}

\abstract{

In order to achieve the challenging requirements on the CLIC vertex detector, a range of technology options have been considered in recent years. One prominent idea is the use of active sensors implemented in a commercial high-voltage CMOS process, capacitively coupled to hybrid pixel readout chips. Recent results have shown the approach to be feasible, though more detailed studies of the performance of such devices, including simulation, are required. The CLICdp collaboration has developed a number of ASICs as part of its vertex detector R\&D programme, and here we present results on the performance of a CCPDv3 active sensor glued to a CLICpix readout chip. Charge collection characteristics and tracking performance have been measured over the full expected angular range of incident particles using 120 GeV/c secondary hadron beams from the CERN SPS. Single hit efficiencies have been observed above \SI{99}{\percent} in the full range of track incidence angles, down to shallow angles. The single hit resolution has also been observed to be stable over this range, with a resolution around 6~$\micron$. The measured charge collection characterstics have been compared to simulations carried out using the Sentaurus TCAD finite-element simulation package combined with circuit simulations and parametrisations of the readout chip response. The simulations have also been successfully used to reproduce electric fields, depletion depths and the current-voltage characteristics of the device, and have been further used to make predictions about future device designs.

}

\titlecomment{This work was carried out in the framework of the CLICdp collaboration}

\begin{document}

%\linenumbers

% generates the title page
\titlepage

\section{Introduction}

%Aim of the paper is to give an overview of the tracking performance achievable, and the variation with angle that can be expected (such that performance over the full angular acceptance of the detector can be gauged). 

The proposal of the high-energy electron-positron Compact LInear Collider (CLIC) at CERN has led to a broad detector R\&D program covering silicon detectors with the aim of providing detectors for the tracker and vertex sub-detectors \cite{CLICCDR_vol2}. The requirements on the vertex detector are particularly stringent and have led to the proposal of the current detector model \cite{Arominski:2649437}, which has three double layers in the barrel region composed of thinned hybrid pixel detectors. These must contain minimal material (of order 50~$\micron$ each for the sensor and readout ASIC) and be able to reach a timing precision of $<$ 5~ns with a single hit resolution of 3~$\micron$. A prototype readout chip (CLICpix) with 25~$\micron$ square pixels has been produced in a 65~nm process technology in order to facilitate the testing of various sensor technologies \cite{clicpix}. Results have previously been presented demonstrating the feasibility of utilising the emerging technology of capacitively coupled pixel detectors using a CCPDv3 sensor fabricated in a 180~nm HV-CMOS process~\cite{Tehrani:2048684}. 

%High single hit efficiencies have been shown for tracks

%The high effciency along with the resolution are paramount for flavour tagging and will aid with achieving CLIC's challenging physics program.

%The angular acceptance of the barrel detector extends to around 75~$^{\circ}$. It is important to characterise the 

The angular acceptance of the current CLIC vertex detector design leads to tracks impinging on the sensor surface at angles of up to \ang{75} with respect to the detector normal. While the performance of capacitively coupled pixel detectors for CLIC has been measured for perpendicular track incidence, a detailed study of the performance over the full angular range is required to ensure their suitability for applications in high energy physics. This work aims to describe this performance, in combination with detailed TCAD simulations of the electric field and charge collection properties of HV-CMOS sensors, and current simulations and parametrisations of the readout response.

\section{Experimental setup}

%A description of the chips, the telescope and how data was taken

\begin{figure}[b!]
	\centering 
	\includegraphics[width=\textwidth]{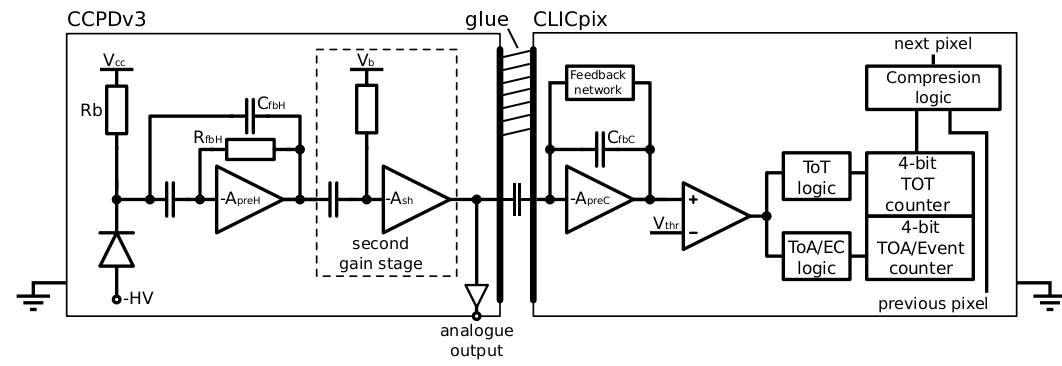}
	\caption{Schematic of the CCPDv3 and CLICpix pixels, with two-stage amplification shown on the CCPDv3.}
	\label{schematic}
\end{figure}

A detailed description of the setup can be found in \cite{Buckland:2633983}. A schematic of the capacitively coupled pixel detector is shown in figure \ref{schematic}, with the CCPDv3 HV-CMOS sensor coupled via a thin layer of glue to the CLICpix readout ASIC. Both pixel chips contain a matrix of 64~$\times$~64 square pixels with 25~$\micron$ pitch, with a capacitance formed between the two by means of pads on their respective top-most metal layers (typically used as the contact for bump-bonding). The pad sizes are of order 20~$\times$~20~$\micron^2$ for the sensor and 14~$\times$~14~$\micron^2$ for the readout ASIC.

The CCPDv3 contains a charge integrating amplifier, with a second gain stage implemented to provide higher output signals. The electronics for each pixel are inside the deep n-well, which acts as the collection diode for the deposited charge. A ring of p-type silicon surrounds each deep n-well allowing the application of a reverse substrate bias of up to 60~V.

The CLICpix chip contains a charge integrating amplifier connected to a discriminator and extended digital logic. Two 4-bit counters record the magnitude and arrival time of the signal, using a Time over Threshold (ToT) and Time of Arrival (ToA) measurement of the amplifier output, respectively. For this study the timing performance of the assembly could not be measured in the beam tests due to the limited dynamic range of the ToA measurement.

Charged hadrons with momentum 120~GeV/c were used to carry out beam tests in the CERN SPS North Area on the H6 beam line. The AIDA telescope \cite{Jansen2016}, composed of 6 planes of MAPS detectors, was employed to supply tracking information and provided a pointing resolution at the Device Under Test (DUT) of 1.6~$\micron$. The DUT was mounted on a precision rotation stage in the centre of the telescope. A region of interest hit counter was used to trigger the DUT frame readout to optimise data taking, given the large difference in area between the DUT and telescope planes (1.6~$\times$~1.6~mm$^{2}$ versus 10~$\times$~20~mm$^{2}$). To correct for threshold dispersion on the readout chip, an equalisation was performed using the 4-bit threshold adjustment on each pixel. In order to be above the noise level, the discriminator threshold voltage ($V_{thr}$) was set to approximately 1000\,e$^{-}$ above the observed baseline output voltage. At this threshold there were found to be around 70 noisy pixels, defined as pixels with a response rate greater than 5$\sigma$ above the matrix mean. This corresponds to less than \SI{2}{\percent} of all pixels. In addition to masking the noisy pixels, a circular mask was also applied to the matrix to account for the variations in the coupling strength introduced during the fabrication process; this is discussed further in section \ref{sec:TrackCharge}. A clock frequency of \SI{20}{\mega\hertz} was used for the ToT measurements.

\section{TCAD simulation}

\begin{figure}[!b]
	\centering 
	\includegraphics[width=.45\textwidth]{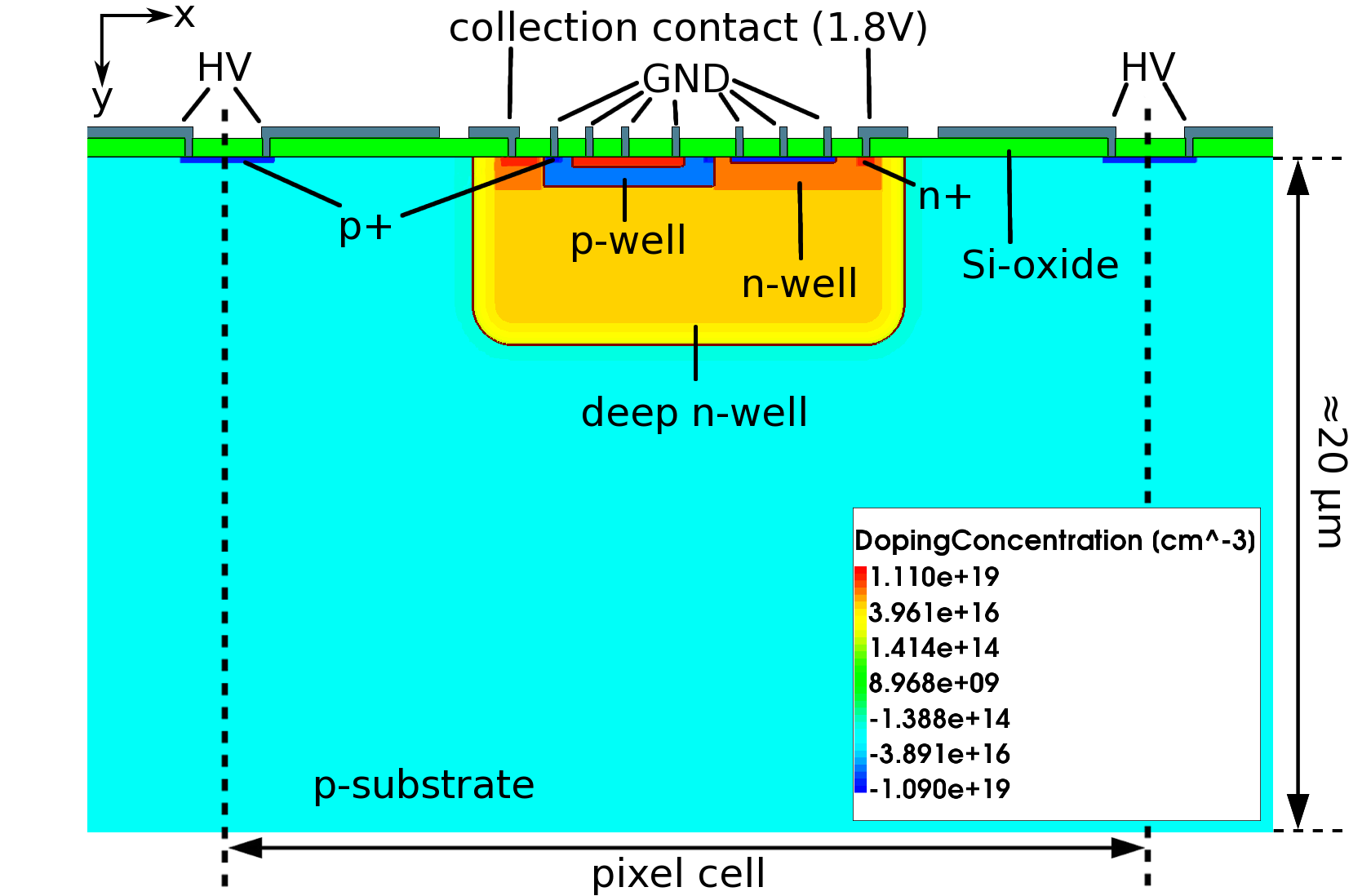}
	\caption{The simulated 2D TCAD structure of the CCPDv3 showing a \SI{25}{\micro\m} pixel cell. The deep n-well contains the implants that make up the in-pixel electronics and is surrounded by a p+ ring where the high voltage is applied.}
	\label{fig: TCAD structure}
\end{figure}

In order to produce the 2D simulation model for use in Synopsys TCAD version I-2013.12 \cite{TCAD}, the implant layout of the CCPDv3 was first extracted from the design GDS (Graphic Data System) file. From this, masks for the different layers were produced and utilised by Sentaurus Structure Editor to build the final structure. The final structure produced by the simulation, for a single pixel cell, is shown in figure \ref{fig: TCAD structure}, indicating the coordinate system used. The bulk silicon is p-type with a resistivity of \SI{10}{\ohm\cm}, with the collection diode consisting of a deep n-well in which the pixel electronics are implanted. The doping profiles used in the simulations are based on typical values found for such a device \cite{TCADdoping}. Table \ref{tab:DopProf} gives the parameters of the doping profiles used in the simulation. A mesh was generated that focused on key areas such as those around the implants, the depletion region and the MIP path. More details on this method can be found in \cite{Buckland:2159671}.

Due to the need for comparison with angled data, the TCAD model was extended to include a row of 10 pixels to better represent the real sensor. The 10-pixel model has a total width of \SI{250}{\micro\m} and a thickness of \SI{250}{\micro\m}, with periodic boundary conditions added to the sides in order to minimise edge effects and best replicate a large sensor. The electrical properties of the CCPDv3 were simulated, in order to compare with measurements of the device performance and provide insights into areas of improvement. 

\begin{table}[!t]
	\begin{center}
	\begin{tabular}{lccc}
		\toprule
		Implant & Doping type & Concentration [\si{\per\cubic\centi\m}] & Gaussian width [\si{\micro\m}]\\ \midrule
		Deep n-well & Phosphorus & \num{1d17} & 1.6\\
		n-well & Phosphorus & \num{1d18} & 0.2\\
		n+ & Phosphorus & \num{1d19} & 0.05\\
		p-substrate & Boron & \num{1.3571d15} & N/A\\
		p-well & Boron & \num{1d18} & 0.2\\
		p+ & Boron & \num{1d19} & 0.05\\
		\bottomrule
	\end{tabular}
	\end{center}
	\caption{Parameters of the simulated doping profiles.}
%	\vspace{5mm}
	\label{tab:DopProf}
\end{table}

Within TCAD the charge deposited by a minimum ionising particle (MIP) was simulated using the Heavy Ion Model from Sentaurus Device. In this model charge is generated uniformly along a path with user-defined parameters, such as position, x- and y-directions. For this study 80 electron-hole pairs per \si{\um} were deposited, without taking Landau fluctuations into account (details of the parameters are given in \cite{Buckland:2633983}). To obtain the current from the collection electrodes, a transient simulation from \SIrange{0}{10}{\micro\s} was performed at bias voltages of \SI{0}{\volt} to \SI{-80}{\volt}. For the simulation of tracks at perpendicular incidence, MIPs were simulated across the central pixel cell in steps of \SI{1}{\micro\m}, while for angled tracks only one MIP position was used at the centre of the second pixel to maximise the track path length through the sensor.

\section{Assembly calibration}

\begin{figure}[!t]
	\centering 
	\includegraphics[width=.45\textwidth]{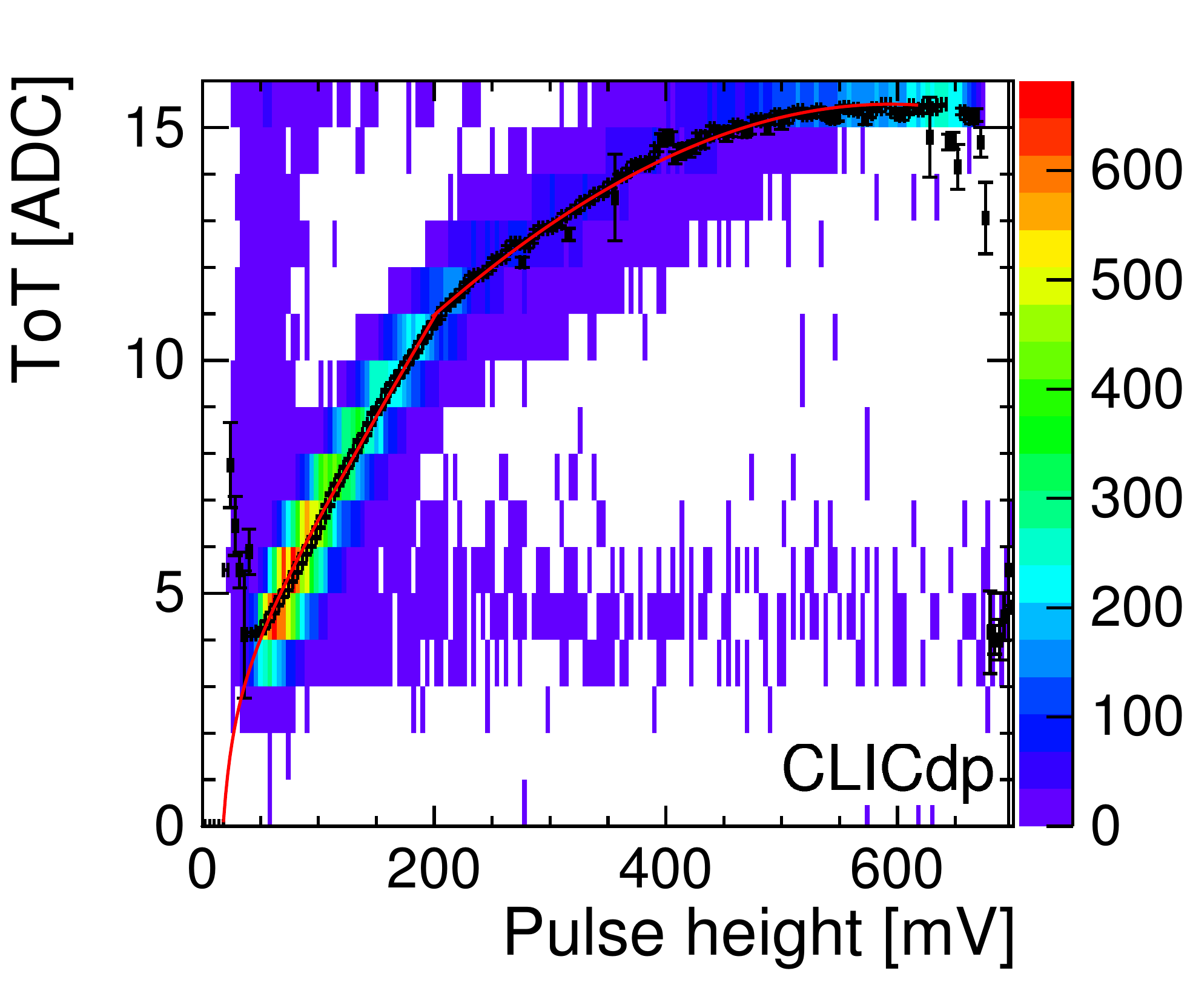}
	\caption{The ToT response as a function of the pulse height from the CCPDv3 amplifier output of a single pixel. The raw data is overlaid with the mean value of the bins represented by the black points. The surrogate fits are shown in red.}
	\label{fig: calibration}
\end{figure}

In order to compare the simulation results to data, a conversion from the simulated signal to an observable ToT measurement on the CLICpix is required. This was performed in two stages: first, the current pulse from TCAD was fed into a circuit simulation of the CCPDv3 electronics. This was carried out using Cadence Virtuoso software \cite{Cadence}, and allowed the analogue output of the CCPDv3 to be obtained. A calibration curve, taken from data, was then used to convert this pulse height into a ToT counter value. The calibrations were conducted in the lab using back-side illumination with a $^{90}$Sr source (with activity \SI{29.6}{\mega\becquerel}) at a bias voltage of \SI{-60}{\volt} to match the beam test conditions. The analogue response of the 2nd stage amplifier within a single CCPDv3 pixel was monitored with a fast sampling oscilloscope and compared to the ToT measurement on the CLICpix. The resulting curve, figure \ref{fig: calibration}, was fit with a surrogate function:

\begin{equation}
\label{eq:SurrFunct}
	\text{ToT} = a\text{P} + b - \frac{c}{\text{P}-t}\,,
\end{equation}

\noindent where $a$, $b$, $c$ and $t$ are the fit parameters and P is the measured pulse height. The low number of data points in the range \SIrange{200}{500}{\milli\volt}, along with the saturation of the ToT, cause problems for the fitting function as there are two separated linear regions. This meant that a satisfactory fit was not possible with just one surrogate function. To address this, the data was fit in two parts, one for the low range data: \SIrange{0}{200}{\milli\volt}, and one for the high range data: \SIrange{200}{620}{\milli\volt}. Mismatches between the pulse height and ToT occur because the oscilloscope and CLICpix have different integration times and their triggers were not synchronised online.

%The points at low pulse height arise from the poor distinction of the signal from the noise resulting in badly fit pulse heights values. The origin of the high ToT, low pulse height points is when a pulse triggers the oscilloscope but not the CLICpix, then while the oscilloscope is waiting, another hit arrives triggering the CLICpix to read out. The points at large pulse heights and low ToT arise because of mismatched double hits. % that produce a large pulse height at varied ToT values.

The CCPDv3 only allows the analogue output to be monitored on 16 pixels equally spaced along the first row, in odd numbered columns. A known issue in the CLICpix front end \cite{Tehrani:2048684,clicpix-twepp-2013}, effecting pixels in the odd set of columns, causes hits to have additional charge injected resulting in a ToT value of less than 2 to increase to this value. The final calibration of energy to a ToT value is obtained by substituting the equation: $Q=CP$ into equation \ref{eq:SurrFunct}, where $Q$ is the charge, $C$ is the test-pulse capacitance equal to the nominal value of \SI{10}{\femto\farad} and $P$ is the pulse height.

\section{Simulation and measurements of capacitively coupled assemblies}

\subsection{Electric field and leakage current}
\label{sec:Efield}

\begin{figure}[t!]
	\centering
	\begin{tabular}{@{}p{0.50\linewidth}@{\quad}p{0.50\linewidth}@{}}
		\subfigimg[width=.45\textwidth]{(a)}{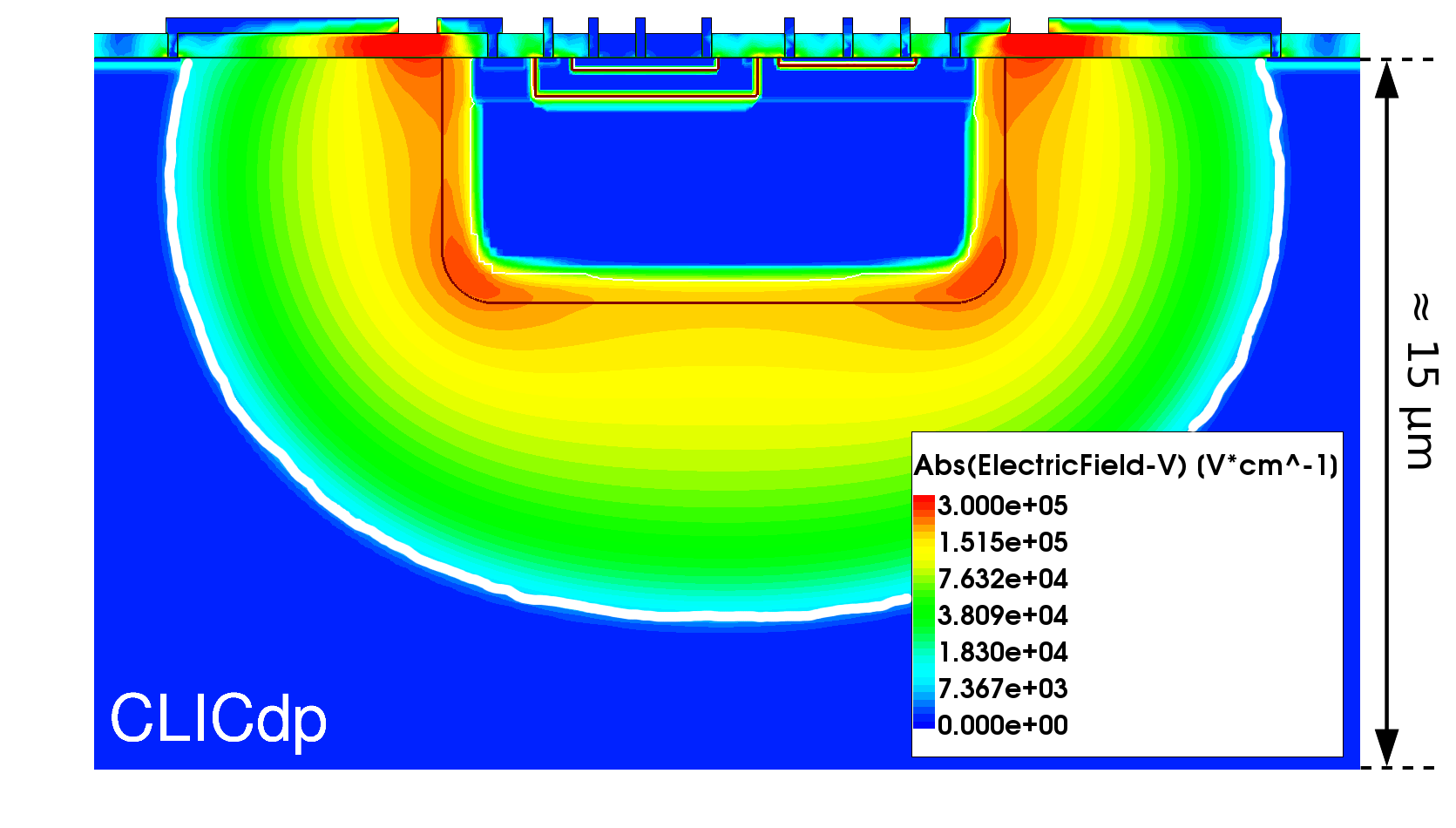} &
		\subfigimg[width=.45\textwidth]{(b)}{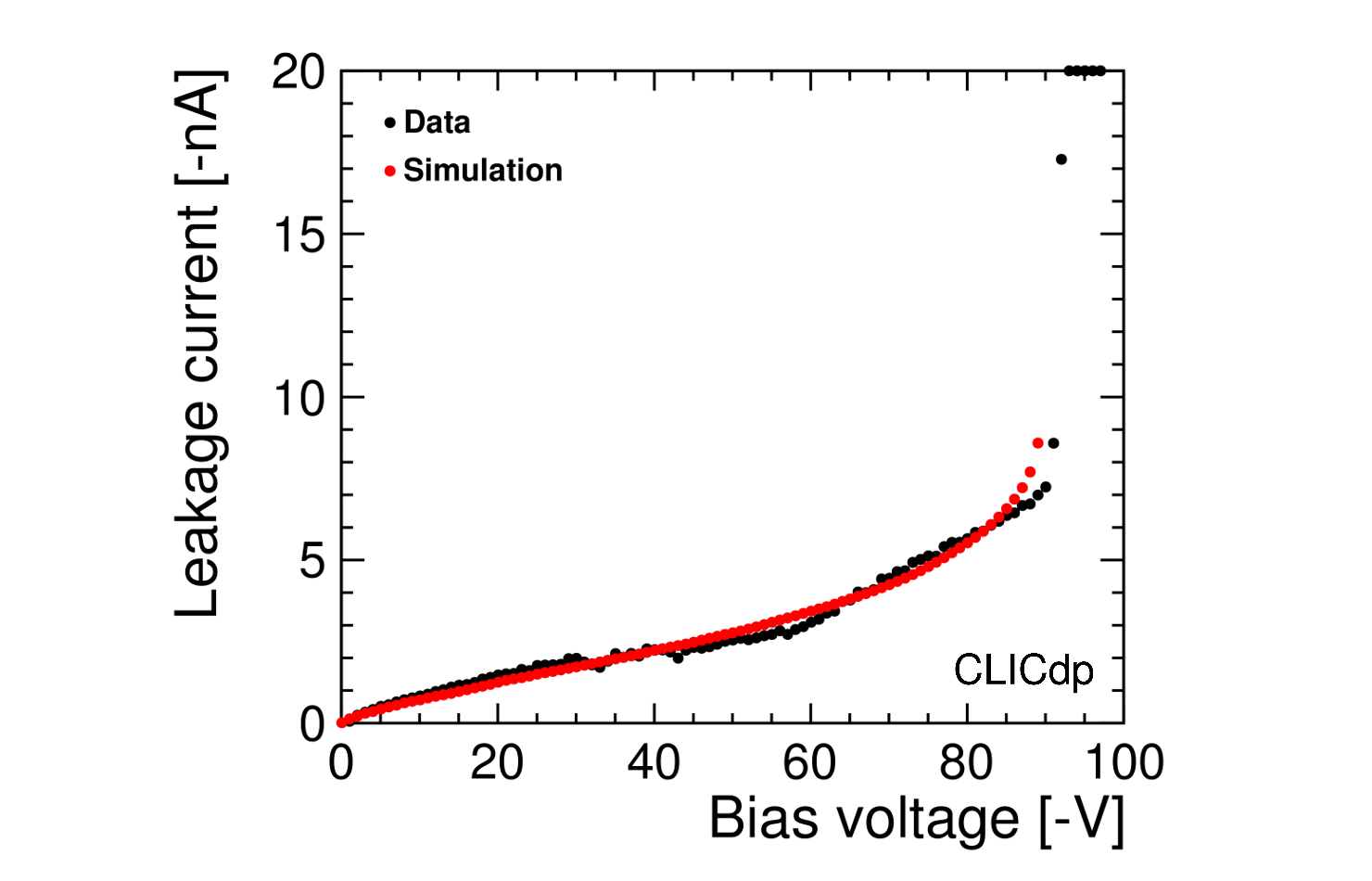}
	\end{tabular}
	\caption{(a) Absolute value of the electric field for a single-pixel cell at the nominal bias of \SI{-60}{\volt}. The white line indicates the border of the depletion region. (b) Current-voltage characteristics for TCAD simulation model (black) and data (red).}
	\label{fig: e-field map}
\end{figure}

The electric field and depletion region of a sensor are important quantities to know as they determine the breakdown characteristics and charge carrier collection properties of the detector. From this, regions within the device that exhibit rapid or delayed charge collection can be identified. A field map showing the absolute value of the electric field at a nominal bias of \SI{-60}{\volt} is shown in figure \ref{fig: e-field map} (a). Over the single-pixel cell the regions of high field are those found around the deep n-well, particularly at the edges, and those in the oxide layer between the HV contact and deep n-well. Outside of the depletion region, the field quickly drops to negligible values. 

The current-voltage characteristics are displayed in figure \ref{fig: e-field map} (b), showing a 
breakdown voltage at \SI{-93}{\volt}.The figure also shows a comparison between simulation and data. In order to scale the 2D simulation to the real detector geometry, the simulated current density per \si{\micro\m} was multiplied by the pixel width (\SI{25}{\micro\m}) and the number of pixels (4096). The current for the simulation and data agree very well, with the largest discrepancy being \SI{5}{\percent}. The breakdown from the simulation is sharper than in the data and occurs at \SI{-88}{\volt}.

\begin{figure}[t!]
	\centering
	\begin{tabular}{@{}p{0.5\linewidth}@{\quad}p{0.5\linewidth}@{}}
		\subfigimg[width=.45\textwidth]{(a)}{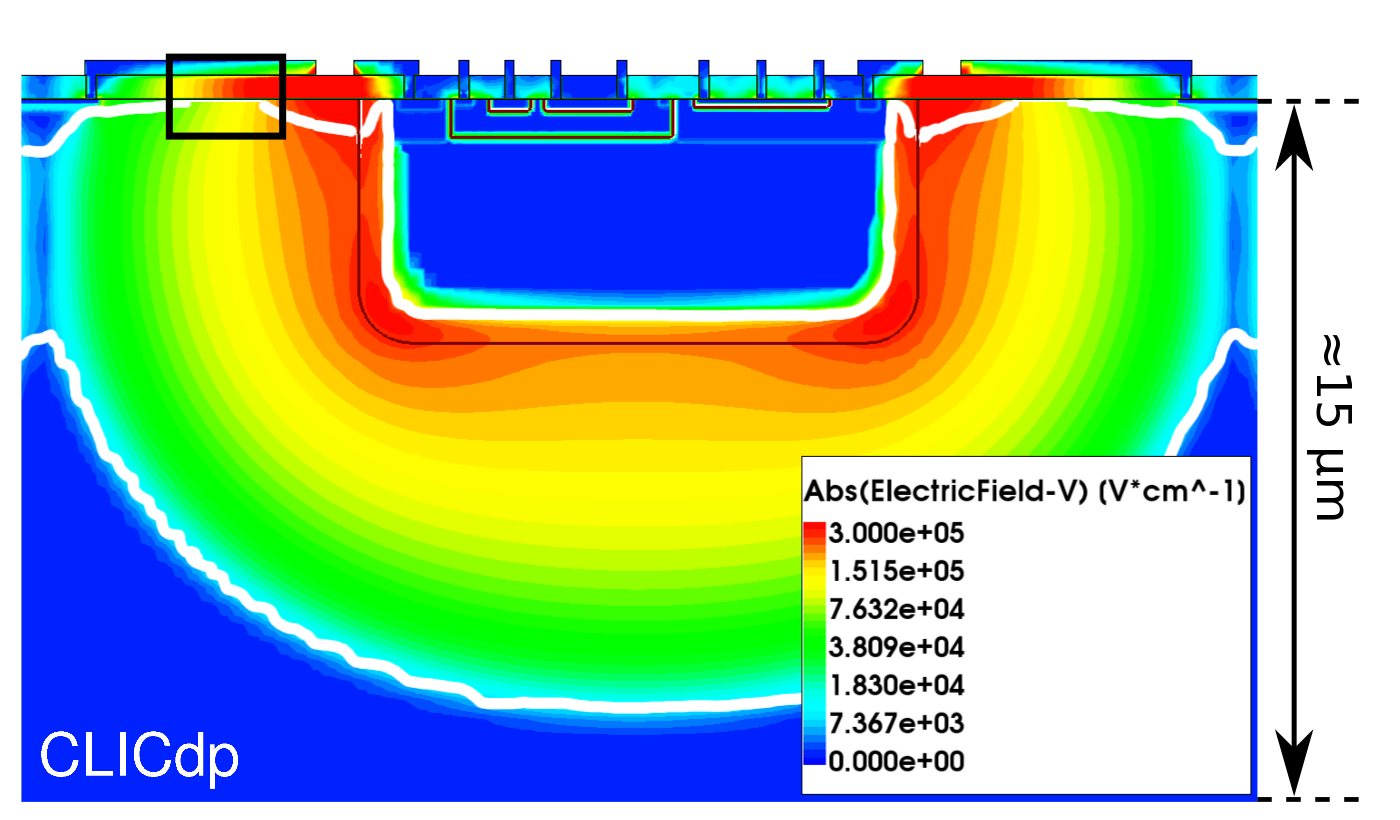} &
		\subfigimg[width=.45\textwidth]{(b)}{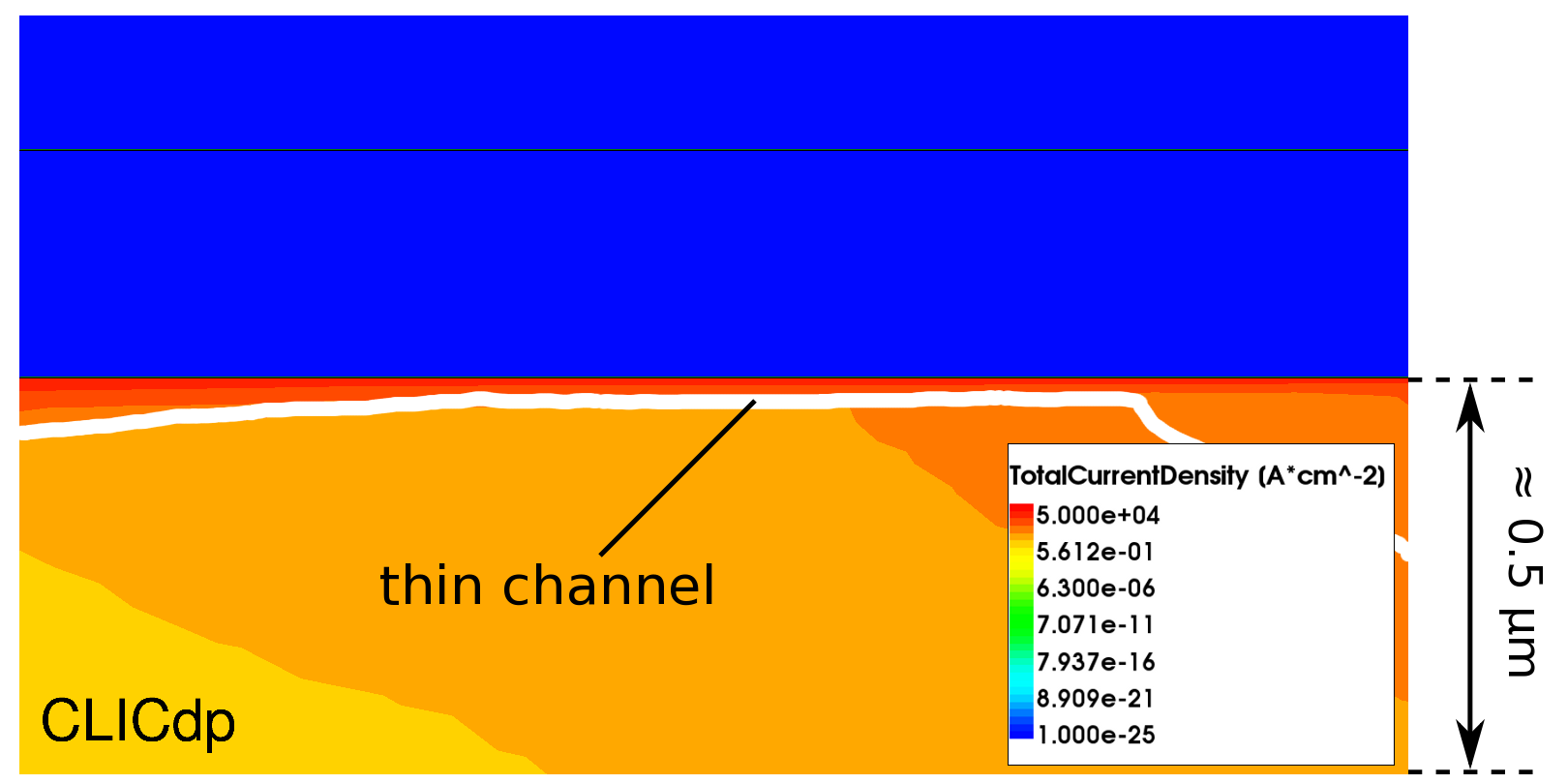}
	\end{tabular}
	\caption{(a) Simulated electric field map at breakdown (\SI{-88}{\volt}). The white line indicates the border of the depletion region. (b) Total current density in the region near the high-voltage contact and edge of the deep n-well, represented by the rectangle on (a).}
	\label{fig: breakdown maps}
\end{figure}

In simulations at the breakdown voltage of \SI{-88}{\volt}, the high field regions in figure \ref{fig: e-field map} (a) become much more pronounced and extend further around the deep n-well (figure \ref{fig: breakdown maps} (a)). In the high field region near the silicon-oxide boundary the electric field is larger than \SI{3d5}{\volt\per\centi\m} (the breakdown field of silicon) and the depletion region becomes distorted, with the production of a thin channel shorting the high-voltage contact and the collection contact. These are the causes of breakdown (sharp rise in current) in the TCAD simulation, highlighted by the large value of the total current density in this region shown in figure \ref{fig: breakdown maps} (b).

\subsection{Depletion depth}
\label{sec:DepDepth}

The sensors do not operating at full depletion, therefore the exact depletion depth is an important quantity to simulate and measure as it is related to the charge collection and timing performance. The depletion width of a p-n junction is proportional to the resistivity of the substrate and the square root of the applied bias voltage. 

\begin{figure}[t!]
	\centering 
	\includegraphics[width=.45\textwidth]{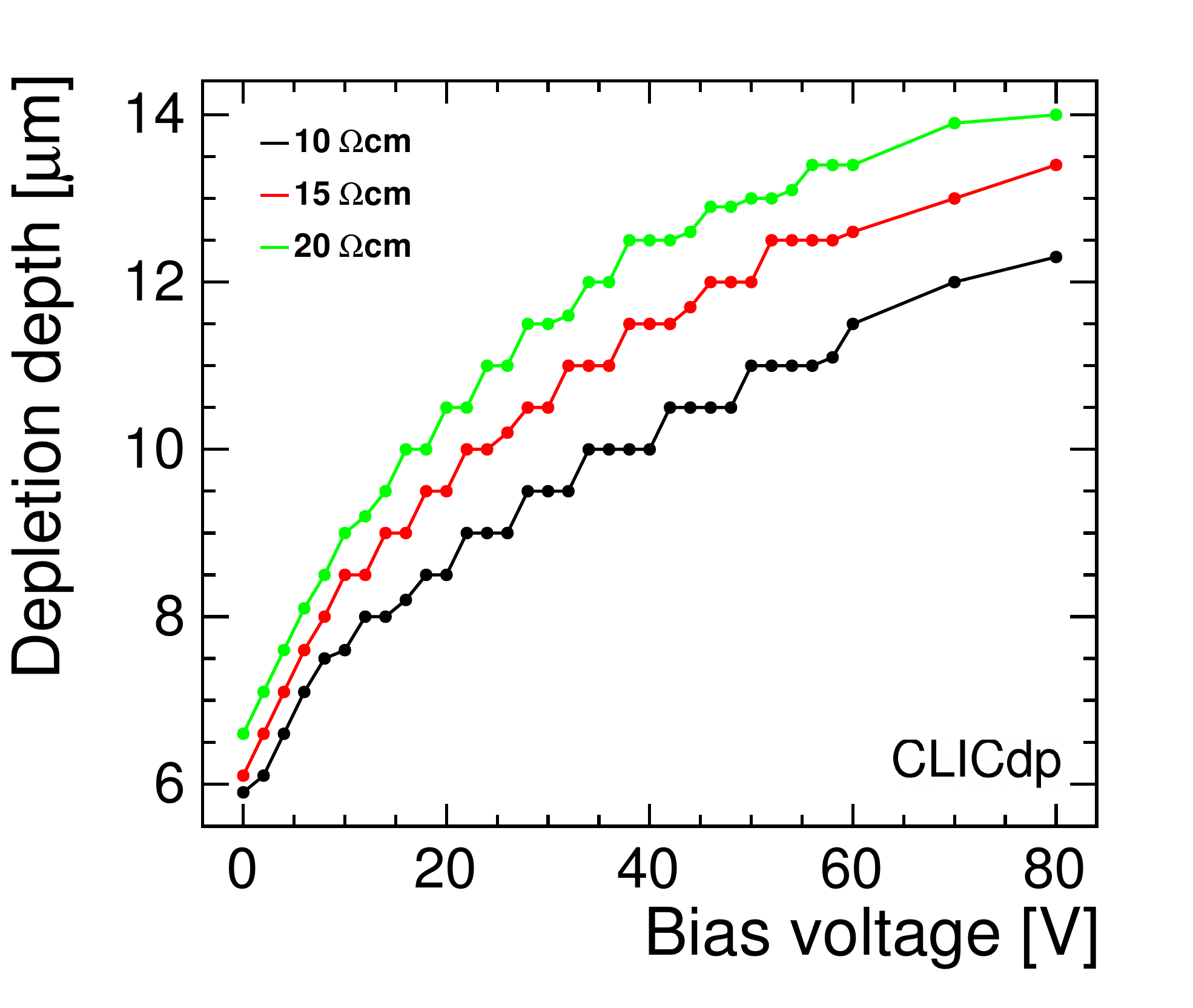}
	\caption{Simulated depletion depth as a function of the applied bias voltage for three different bulk resistivity values.}
	\label{fig: depDepth}
\end{figure}

Measurements using an Edge Transient Current Technique, based on a Two Photon Absorption process (TPA-eTCT), of the depletion depth from the sensor surface at \SI{-80}{\volt} give a value of (15$\pm$1)\,\si{\micro\m} \cite{GARCIA201769}. The measured value of the bulk resistivity of the device is (14.7$\pm$1.7)\,\si{\ohm\cm} \cite{GARCIA201769}, being larger than the nominal value in the simulation of \SI{10}{\ohm\cm}. Therefore, the depletion depth, from the surface of the sensor, obtained from TCAD simulations at \SI{15}{\ohm\cm} and \SI{20}{\ohm\cm} are shown along with the nominal value in figure \ref{fig: depDepth}. The step structure in the plot is an artefact of the simulation mesh and the extraction of the depletion depth. Comparing the measured value to the \SI{15}{\ohm\cm} simulated value of \SI{13.4}{\micro\m} shows a reasonable agreement between the two. This difference of $\sim$12\% between the measured and simulated values could arise from the limitations of the simulation such as it only being in 2D and the simulation being an idealised case with respect to the doping profiles. 

\subsection{Simulated charge collection}
\label{sec:SimCharge}

\begin{figure}[b!]
	\centering
	\begin{tabular}{@{}p{0.5\linewidth}@{\quad}p{0.5\linewidth}@{}}
		\subfigimg[width=.45\textwidth]{(a)}{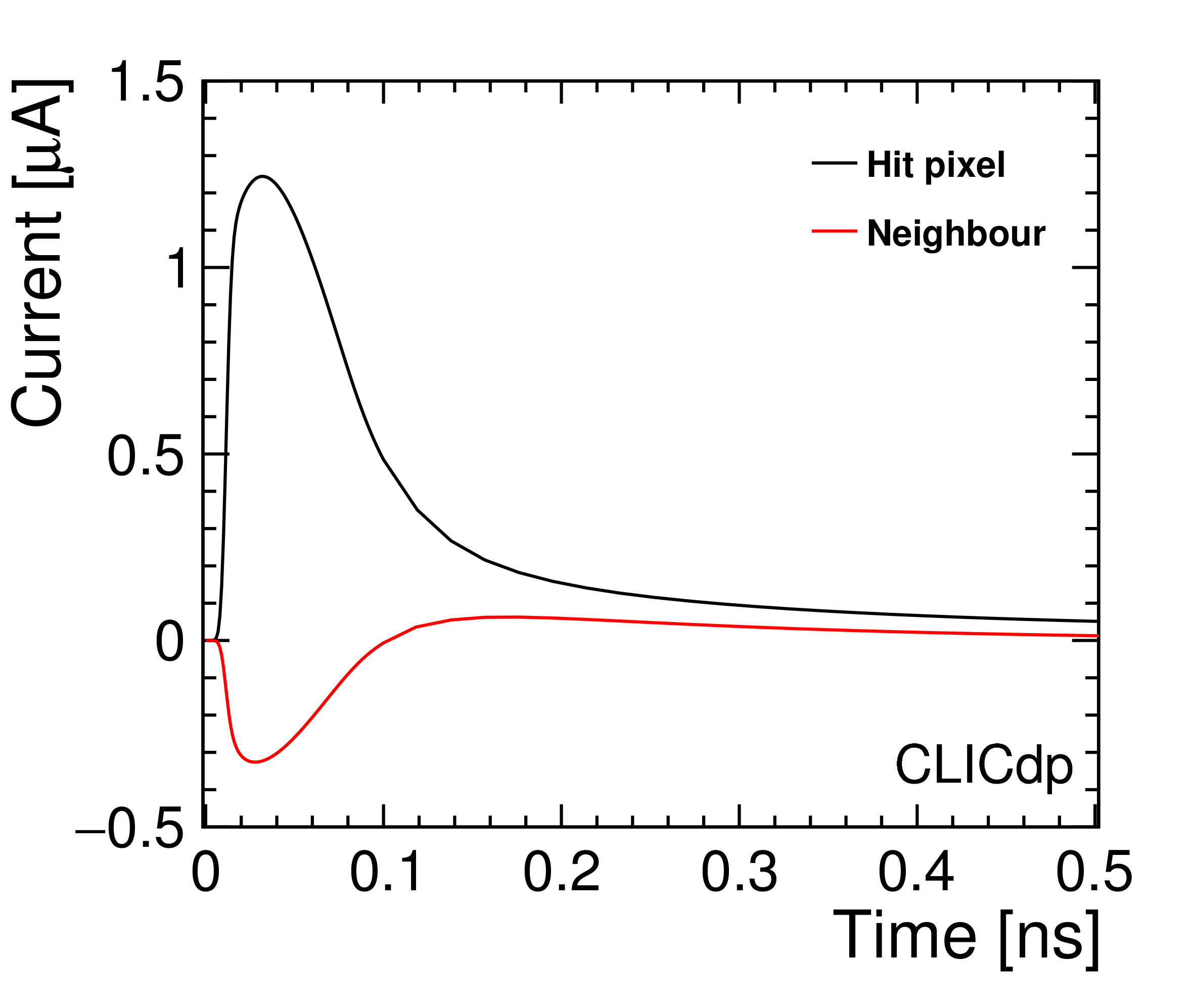} &
		\subfigimg[width=.45\textwidth]{(b)}{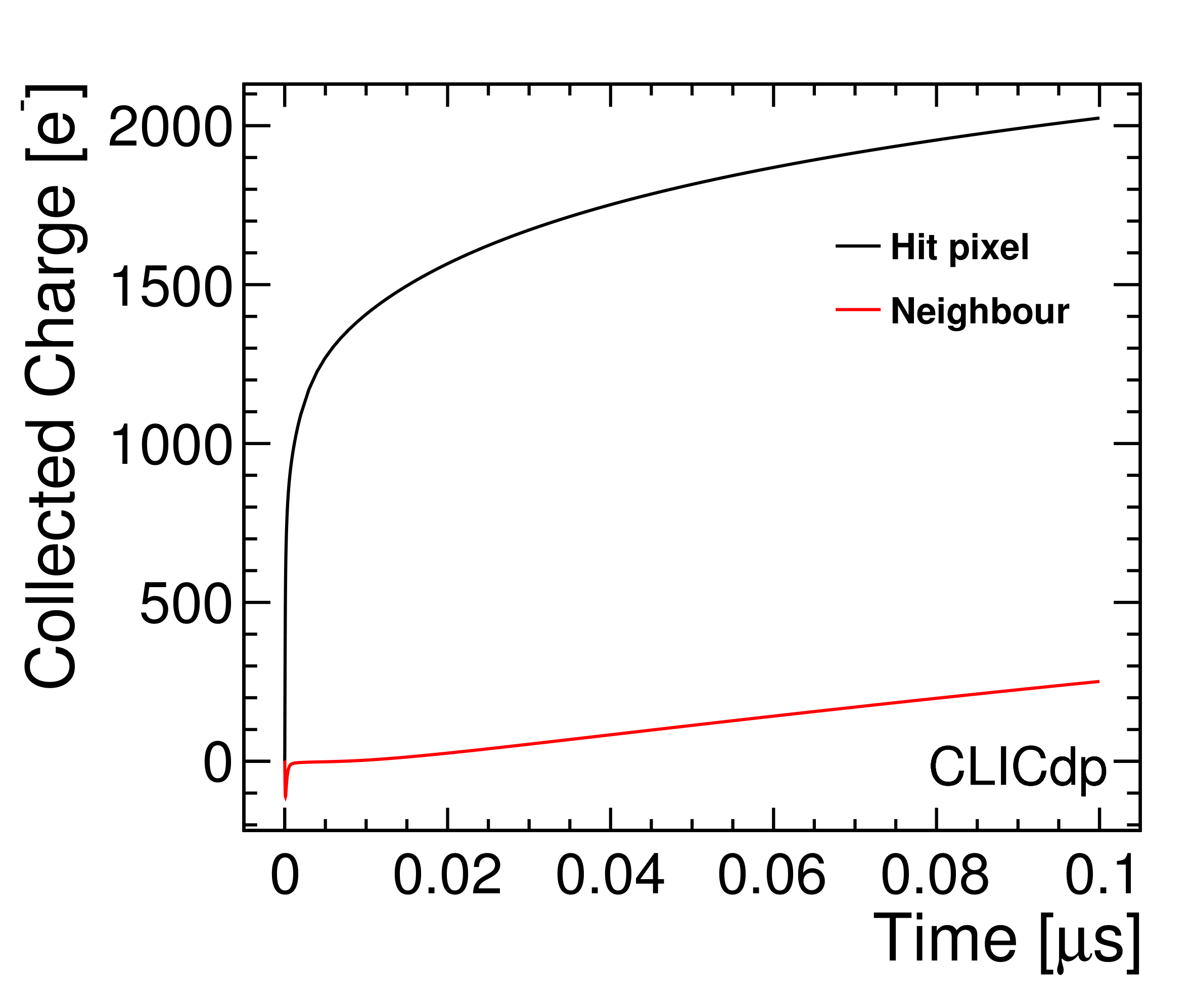}
	\end{tabular}
	\caption{Simulated (a) current pulse and (b) charge collection from a MIP-like charge deposition, for the hit pixel and a neighbouring pixel.}
	\label{fig: simCurrPulse}
\end{figure}

The simulated current pulses from a hit pixel and its adjacent neighbours can be seen in figure \ref{fig: simCurrPulse} (a), for a MIP traversing perpendicular to the surface and passing through the centre of the pixel cell. The peak of the current pulse for the hit pixel occurs very early, at around \SI{25}{\pico\s}, and rapidly reduces to much lower current values beyond \SI{200}{\pico\s}. The neighbouring pixels have, due to the symmetry of the device, identical pulse shapes which are initially negative. This is due to the electrons deposited in the depletion region of the hit pixel moving away from the neighbours, inducing a negative charge in accordance with the Shockley-Ramo theorem. In both cases, the current value is above 0 indicating there is still charge being collected. The integrated charge after \SI{100}{\nano\s} is shown in figure \ref{fig: simCurrPulse} (b), where it can be seen that the neighbour pixel charge is $\approx$ \SI{12.5}{\percent} of the hit pixel.

\subsubsection{Charge collection modes}
\label{sec:CCModes}

\begin{figure}[t!]
	\centering
	\begin{tabular}{@{}p{0.5\linewidth}@{\quad}p{0.5\linewidth}@{}}
		\subfigimg[width=.45\textwidth]{(a)}{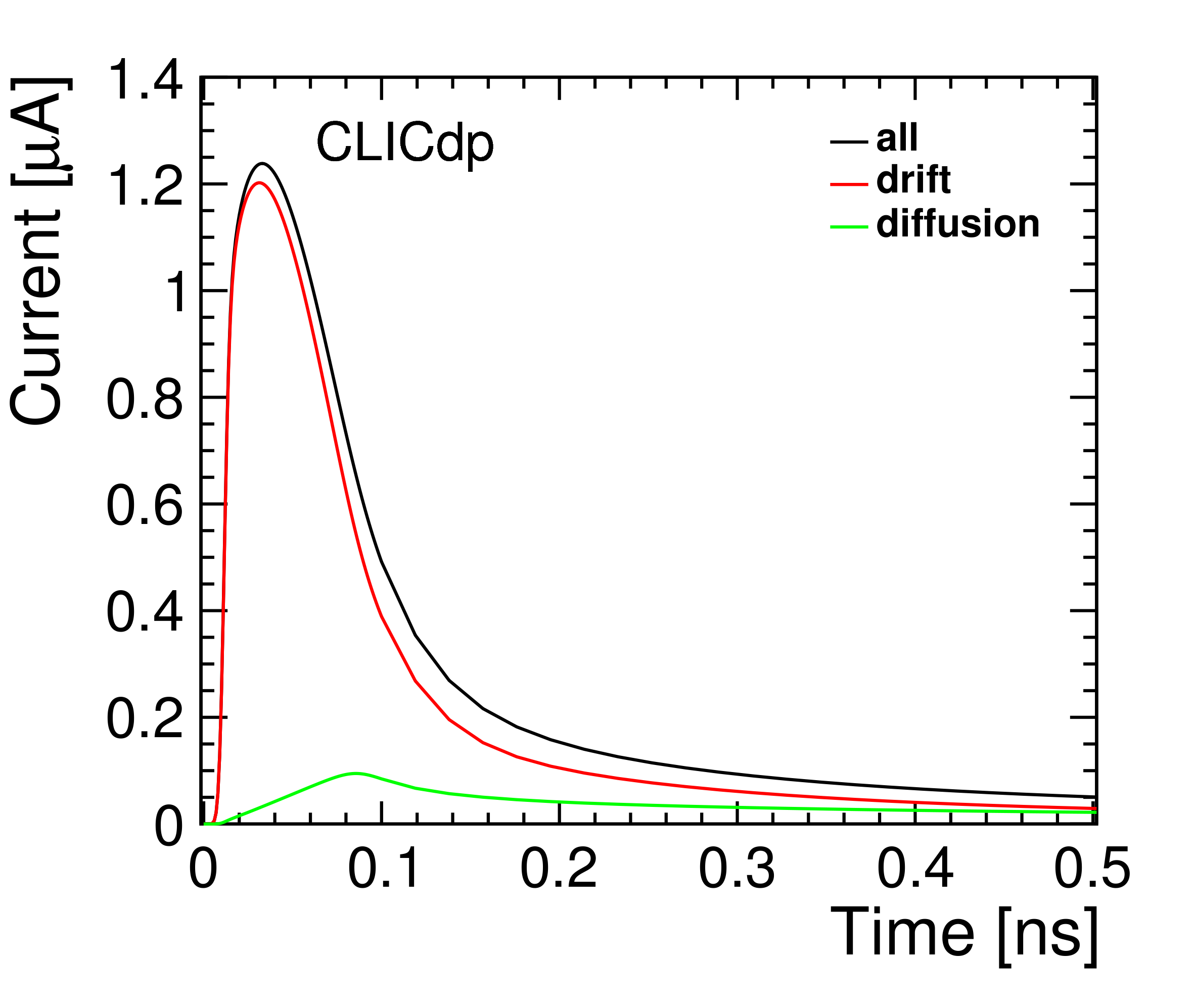} &
		\subfigimg[width=.45\textwidth]{(b)}{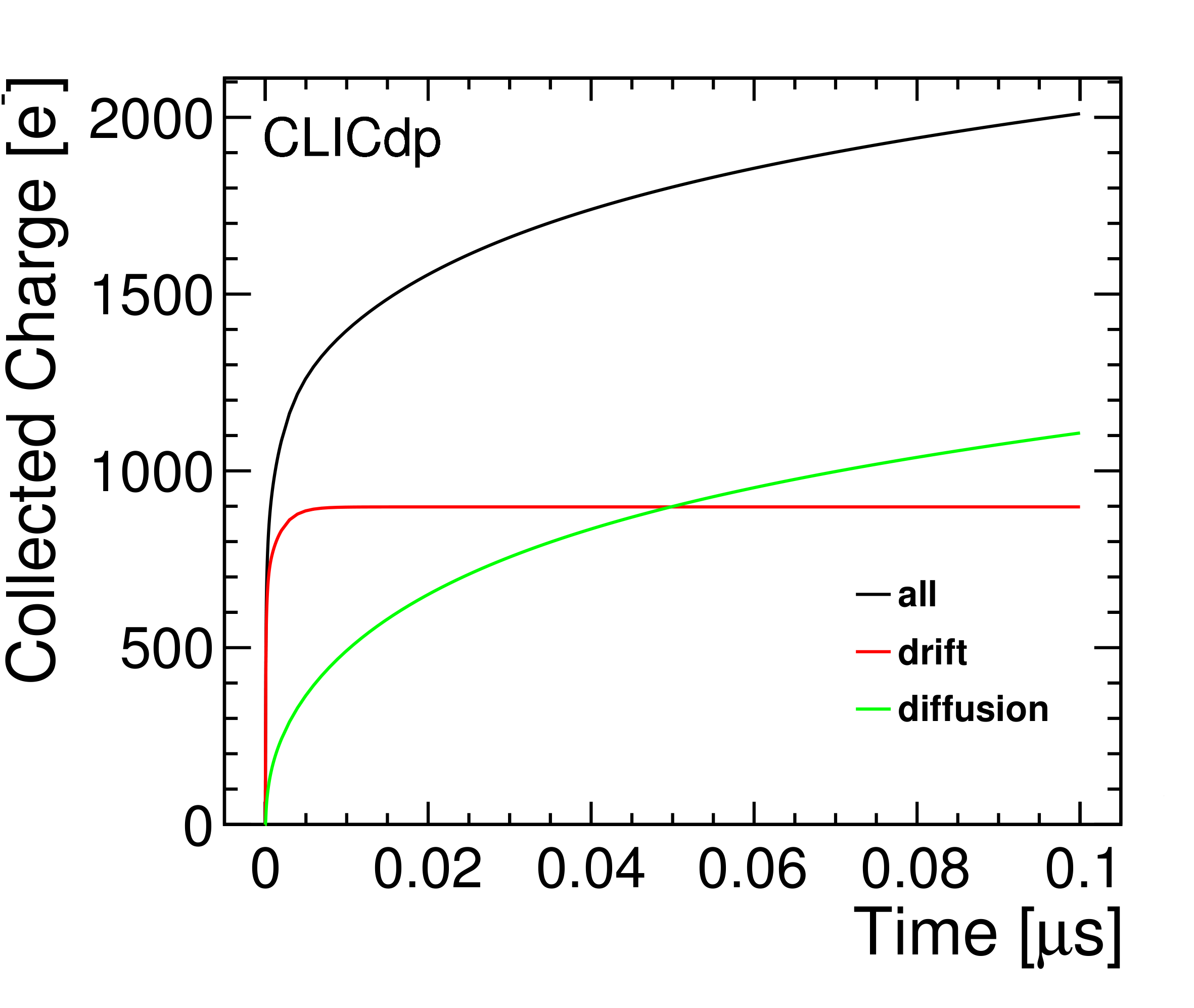}
	\end{tabular}
	\caption{Simulated (a) current pulse and (b) charge collection profile for different charge collecting regions.}
	\label{fig: split MIP}
\end{figure}

As HV-CMOS detectors can have contributions to the collected charge from both inside the depletion region (drift) and outside the depletion region (diffusion then drift), it is interesting to look at the characteristics of both signal pulses on the overall sensor response. To aid this, the charge deposition from the simulated MIP is spit into \textit{drift} and \textit{diffusion} regions, the drift region being defined by the depletion region. In the \textit{drift} case, charge is deposited in the simulation only within the depleted region of the sensor, while in the \textit{diffusion} case charge is deposited only in the non-depleted bulk. The resulting pulse shapes are shown in figure \ref{fig: split MIP}, with both the current pulse and cumulative charge distributions. It can be seen that the full current pulse is dominated by the contribution from charge deposited within the depleted region, and that this charge is fully collected within around \SI{5}{\nano\s}. However, it is interesting to note that after \SI{50}{\ns} the cumulative charge collected by the implant has equal contributions from both drift charge and that from the non-depleted region. This is larger than the integration time of the CLICpix readout chip which is $\sim$\SI{30}{\nano\s} \cite{clicpix}, so collection by drift is dominant.

\section{Tracking Performance}
\label{sec: TrackPerform}

Tracks reconstructed in the telescope only include the telescope planes and are projected onto the DUT. A cluster of pixels is defined as the ToT-weighted centre of gravity and a cluster with energy deposition above the DUT threshold is associated to the track if it lies within \SI{100}{\micro\m} of the track intercept. The track is excluded from further analysis if the intercept lies in a pixel masked during data-taking or in one of its neighbours, or if another track intercepts the DUT within a \SI{125}{\micro\m} radius. In addition, a cut on the track fitted $\chi^{2}$/NDF of 3.5 is applied to the tracks in order to remove those with significant scattering.

\subsection{Charge collection}
\label{sec:TrackCharge}

\begin{figure}[t!]
	\centering
	\begin{tabular}{@{}p{0.5\linewidth}@{\quad}p{0.5\linewidth}@{}}
		\subfigimg[width=.465\textwidth]{(a)}{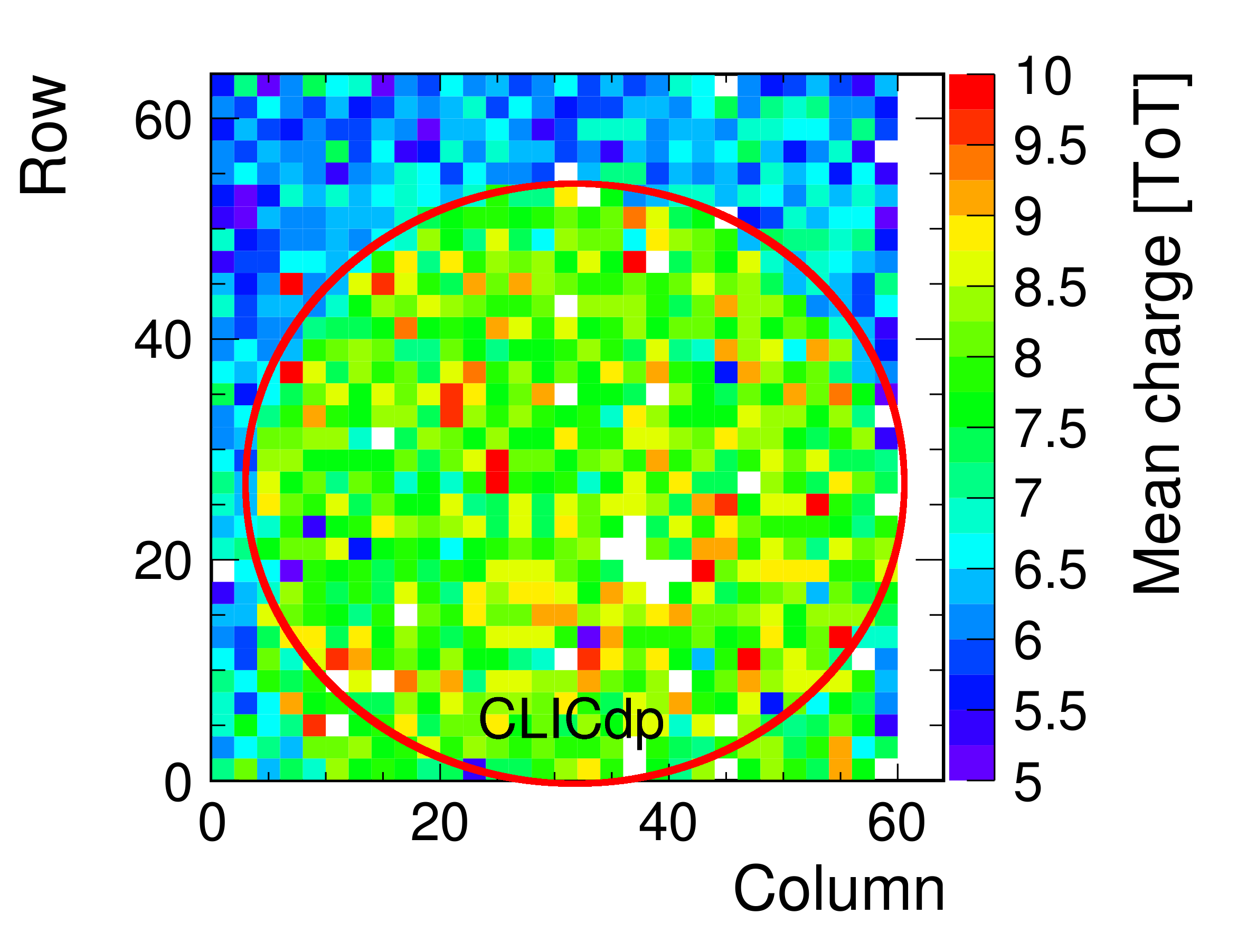} &
		\subfigimg[width=.42\textwidth]{(b)}{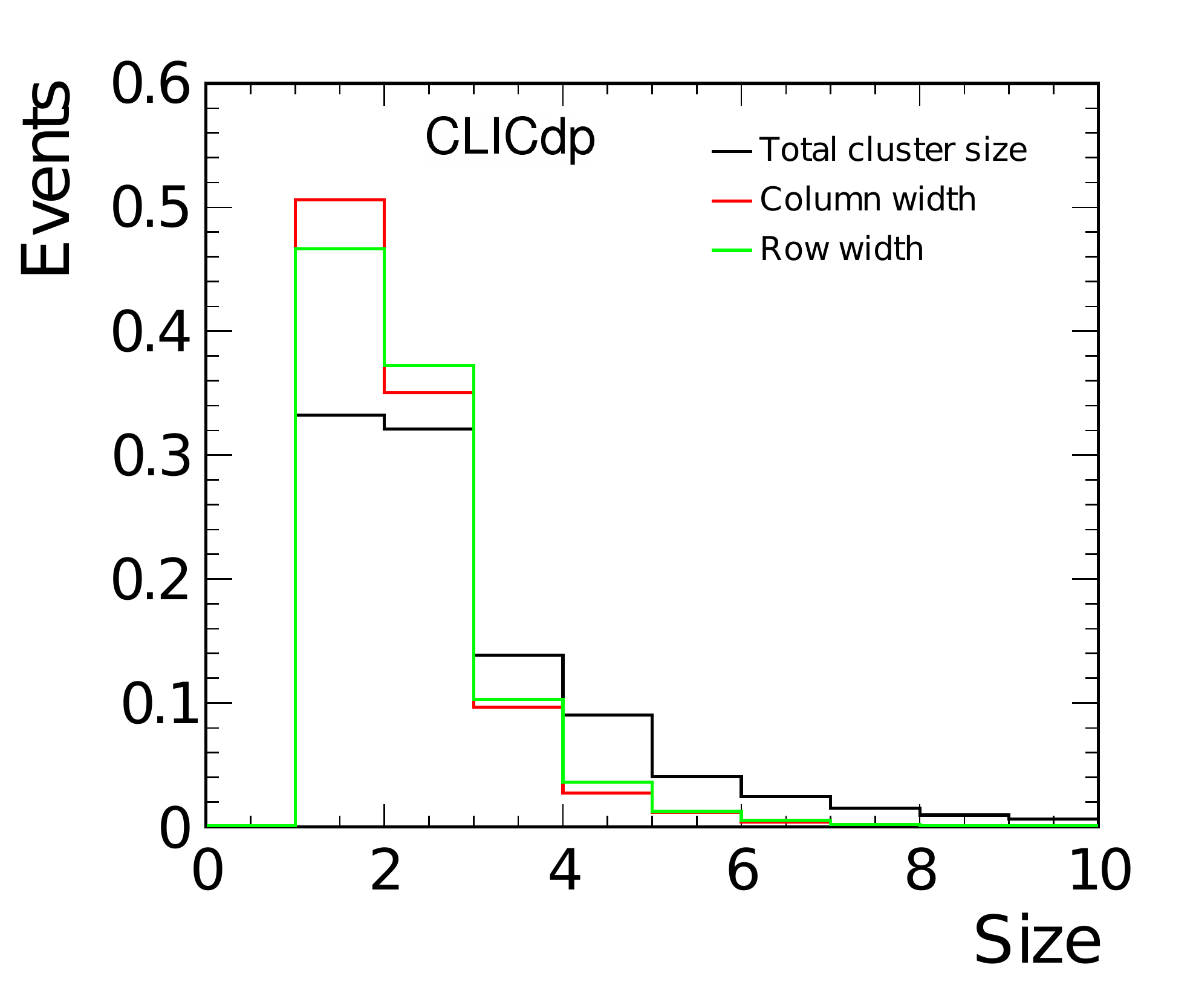}
	\end{tabular}
	\caption{(a) Mean ToT per pixel averaged over 2x2 pixels for DUT hits assigned to telescope tracks at normal incidence. For each pixel position the ToT is fitted with a Gaussian to obtain the mean. The red line indicates the circular mask used. (b) The cluster size distribution, broken down into column and row width.}
	\label{fig: ToT MPV response TCAD}
\end{figure}

In order to compare the simulation output with data, effects of the detector assembly have to be taken into account. The uniformity of the signal across the CLICpix depends on the coupling to the HV-CMOS sensor; the response across the matrix is shown in figure \ref{fig: ToT MPV response TCAD} (a). For each pixel the ToT spectrum is fitted with a Gaussian distribution from which the mean is extracted and plotted in the figure averaged over 2x2 pixels. The choice of fit was chosen because the ToT spectrum does not display the typical Landau shape due to the small resolution of the ToT counter (4-bit). As can be seen, the matrix contains a clearly visible circle in which the pixels have a higher observed signal compared to the rest of the matrix. This is attributed to the glue, which has a significantly larger dielectric constant than air, covering only this circle. The region of lower capacitance has been removed from further analysis by applying a circular mask indicated by the red line in figure \ref{fig: ToT MPV response TCAD}. The cluster size distribution at perpendicular incidence, shown in figure \ref{fig: ToT MPV response TCAD} (b), shows there are mainly 1-2 hit clusters.

\subsection{Cross-coupling}

\begin{figure}[b!]
	\centering
	\begin{tabular}{@{}p{0.5\linewidth}@{\quad}p{0.5\linewidth}@{}}
		\subfigimg[width=.45\textwidth]{(a)}{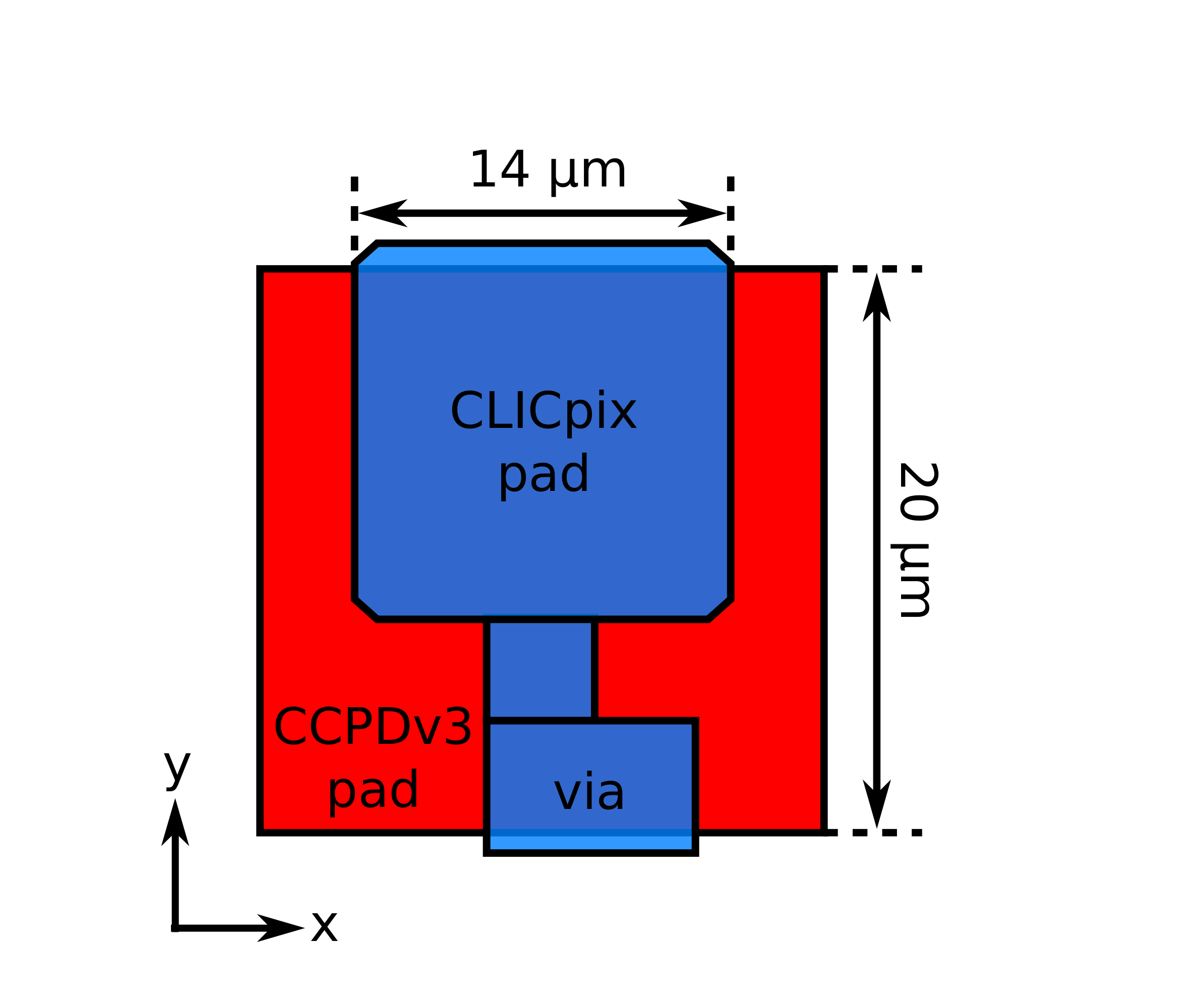} &
		\subfigimg[width=.45\textwidth]{(b)}{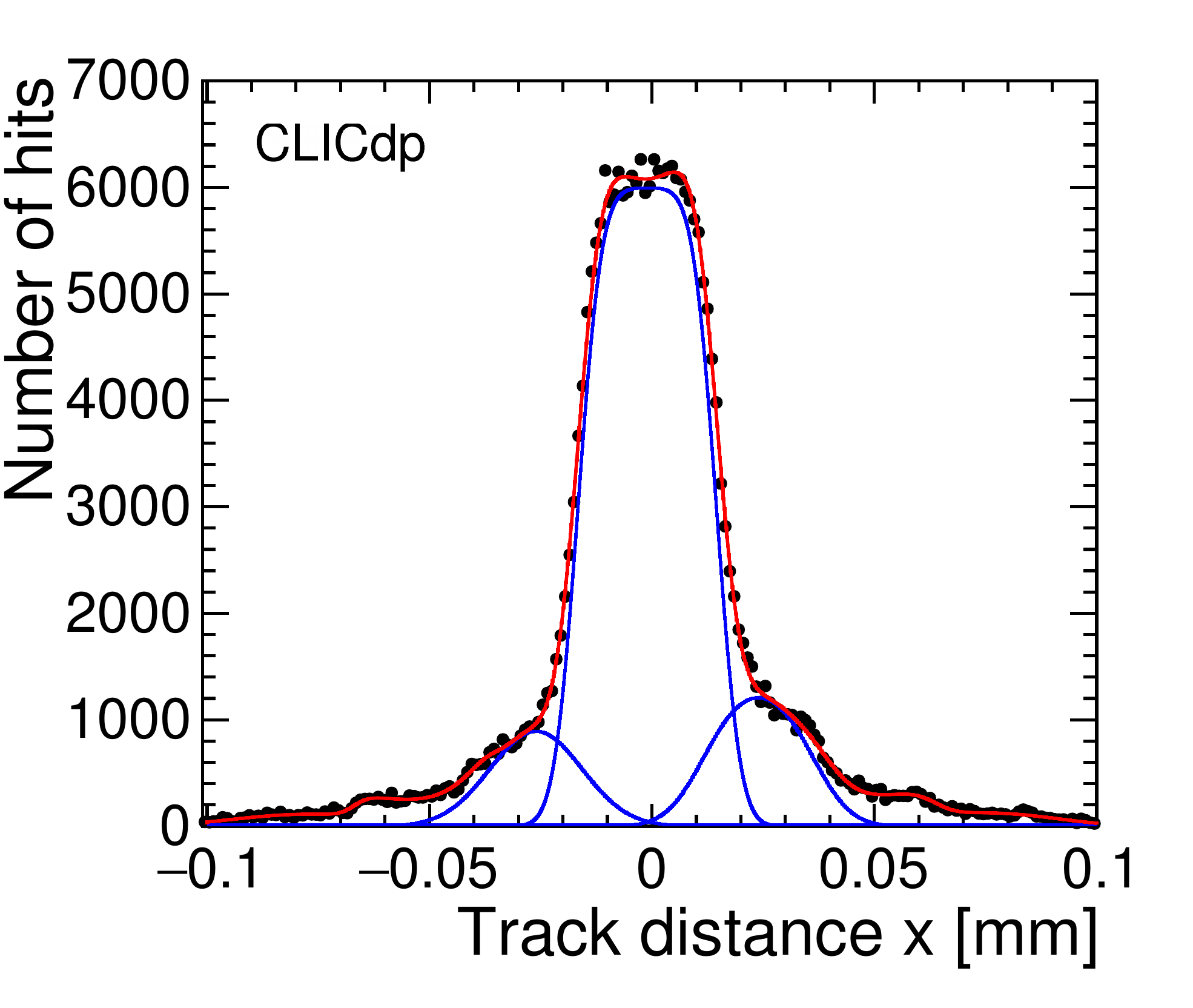}
	\end{tabular}
	\caption{(a) Illustration of the CCPDv3 and CLICpix coupling pads, including the via located on the CLICpix. (b) Pixel response as a function of distance of the track position from the pixel centre at perpendicular incidence. The total top hat fit is in red while individual top hat fits for three pixels are in blue, detailed in \cite{Tehrani:2048684}.}
	\label{fig: crossCoup}
\end{figure}

In addition to the uniformity of the pixel matrix response, the relative alignment of the HV-CMOS and readout ASICs is an important factor in the measured performance. The coupling pads through which the signal is transferred from the CCPDv3 to the CLICpix are shown in figure \ref{fig: crossCoup} (a), in their nominal alignment and orientation for bonding. As the signal is transferred capacitively and the pads are large compared to the inter-pixel distance, capacitive coupling from a single HV-CMOS pixel to multiple pixels on the readout chip may take place; this has already been observed for these devices \cite{Tehrani:2048684}. Hits due to cross-coupling can be observed by plotting the pixel response as a function of distance of the track position from the pixel centre at a perpendicular incidence, as per figure \ref{fig: crossCoup} (b). The central peak is due to charge deposited inside the active volume of the pixel and contributions from diffusion to the neighbours, while additional peaks outside of this range are induced by cross-capacitances to neighbouring pixels which also fire. Due to the design of the CCPDv3, cross-talk to neighbours is very small so the additional peaks are due to cross-coupling. The plot has been fitted with a combination of error functions shown by the red line and the contributions from each pixel to the overall fit are given by the blue lines. Details of the fit method are given in \cite{Tehrani:2048684}. The small asymmetry in the pixel response can be explained by asymmetric capacitive cross talk due to misalignment during production of the glue assembly ($\sim$ \SI{2}{\micro\m} alignment precision) \cite{Tehrani:2048684}.

\begin{figure}[b!]
	\centering 
	\begin{tabular}{@{}p{0.5\linewidth}@{\quad}p{0.5\linewidth}@{}}
		\subfigimg[width=.45\textwidth]{(a)}{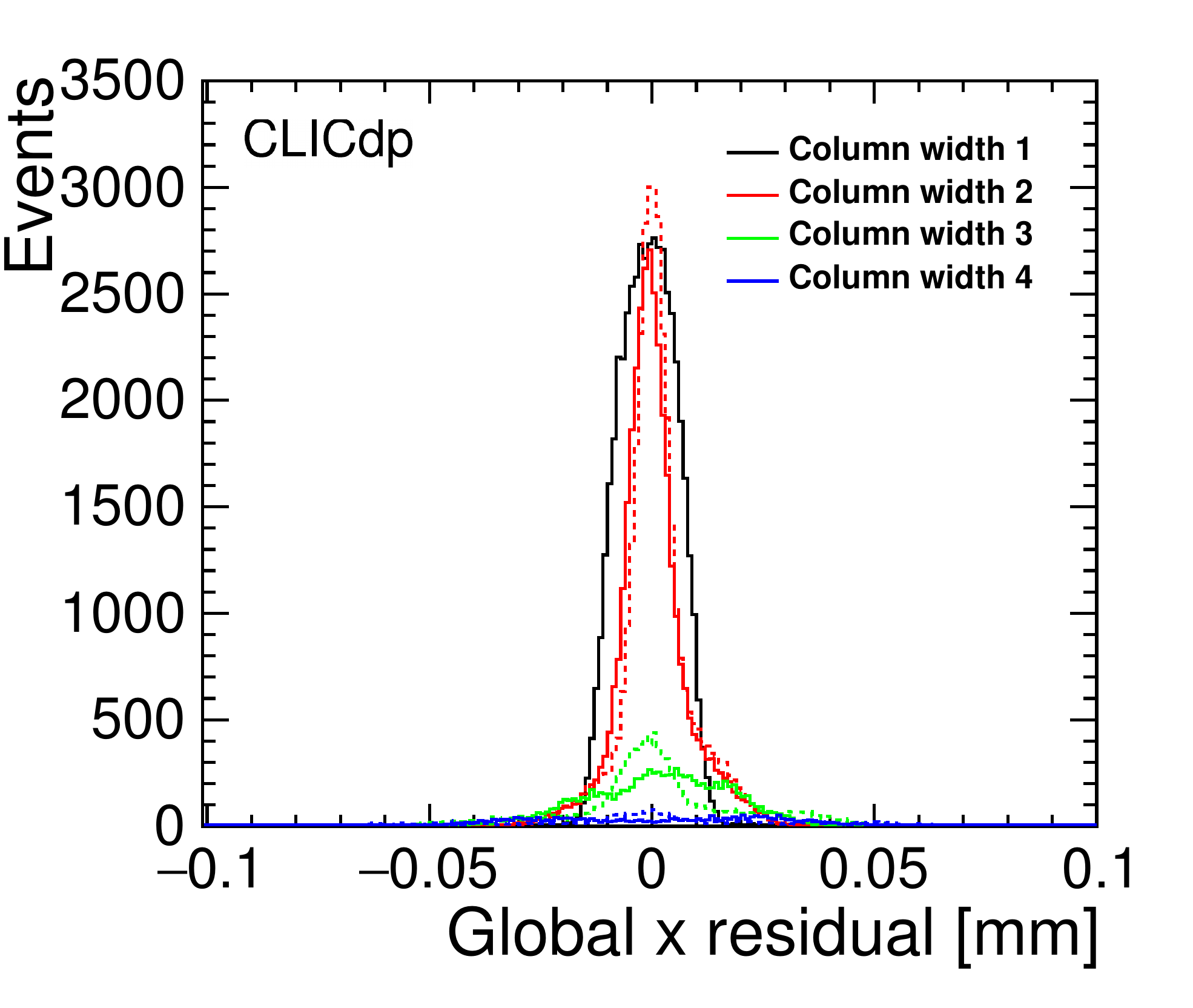} &
		\subfigimg[width=.45\textwidth]{(b)}{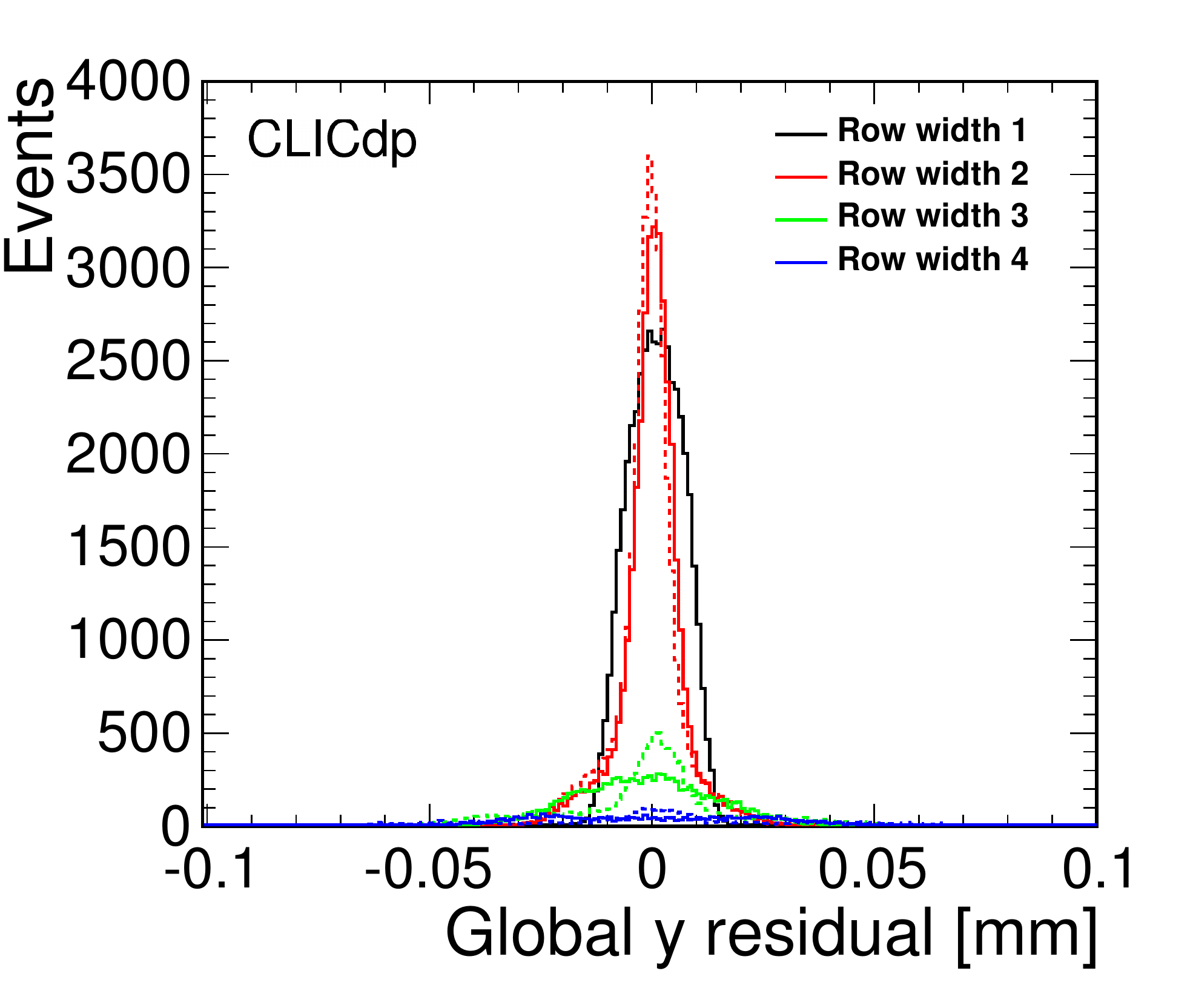}
	\end{tabular}
	\caption{Residuals in the (a) x- and (b) y-direction and at perpendicular incidence for different cluster widths before (solid line) and after eta correction (dashed line).}
	\label{fig: x residual}
\end{figure}

Figure \ref{fig: x residual} shows the individual residual distributions for perpendicular track incidence, both before and after the application of corrections to account for the non-linear sharing of charge between pixels, the so-called eta corrections. For each cluster and separately for widths 2, 3, 4, and 5 and above in the column (x-direction) and row (y-direction) direction, the position obtained from the centre of gravity algorithm was plotted against the track position. The resulting distribution was fitted with a fifth order polynomial, and was subsequently used as the correction factor. While the residuals improve sharply as expected, the effects of multi-hit clusters due to capacitive cross-coupling can be clearly seen. When considering clusters with two pixels (especially after correction), two distinct distributions can be seen: those which are the result of tracks close to the pixel boundaries and hence the charge in each pixel is proportional to the track position; and those where the second pixel has fired due to cross-coupling. In the latter case, the position resolution of the cluster can only be degraded, and a much wider residual distribution, even than for single pixel clusters, is observed. The asymmetry seen in the multi-pixel residuals is a result of the glueing alignment precision as previously mentioned.

\begin{figure}[t!]
	\centering 
		\includegraphics[width=.45\textwidth]{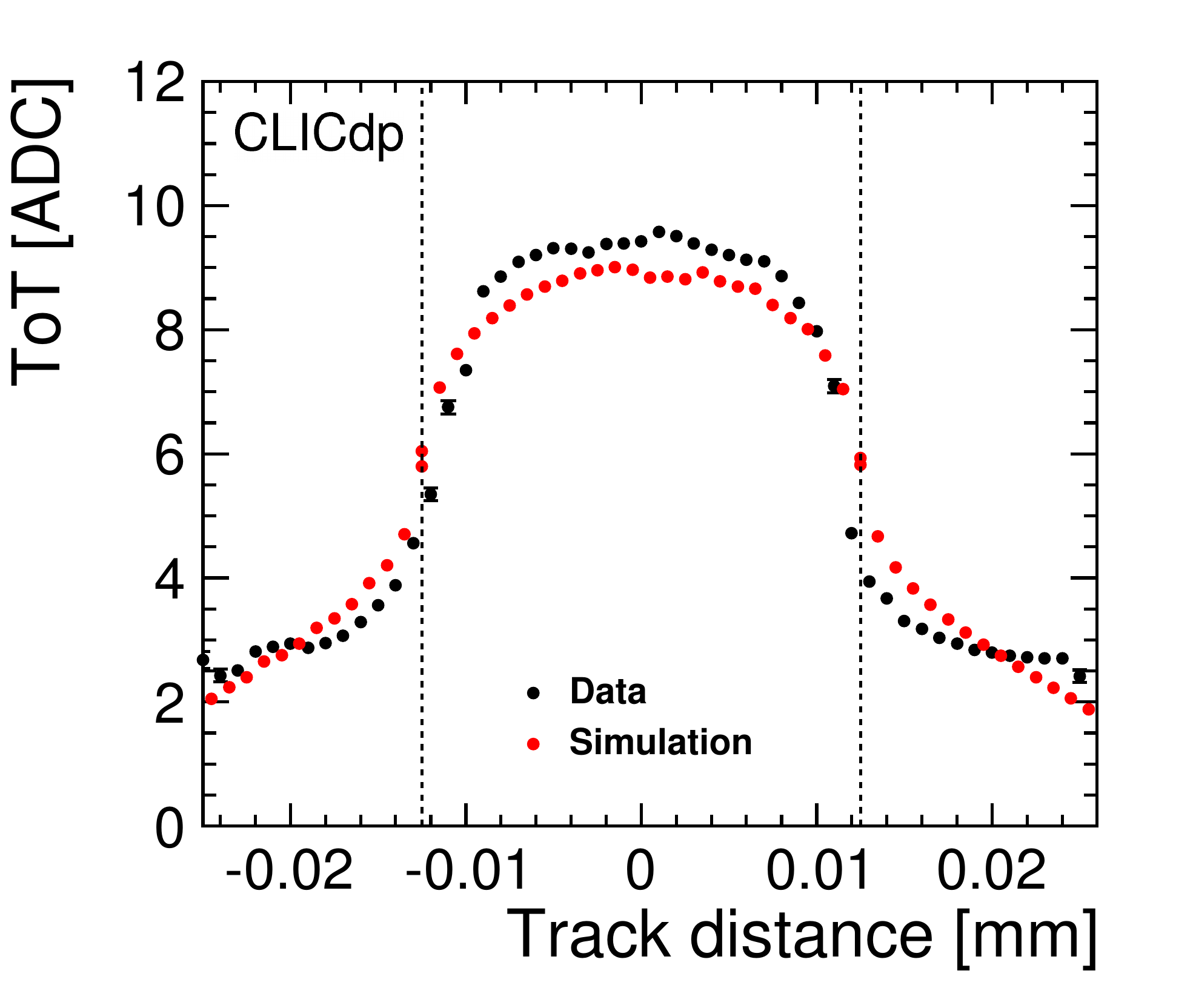}
	\caption{Pixel ToT as a function of track distance in the column direction, for data (black) and simulation (red). The vertical dashed lines represent a one pixel unit cell.}
	\label{fig: pixelToTResponse}
\end{figure}

Since the effect of the cross-coupling is not negligible, an estimate of the cross-coupling capcitance was needed so it could be added to the simulations. This was done using the finite element analysis software COMSOL Multiphysics. The largest capacitance to a neighbour pixel, relative to the main coupling, for an ideal pad alignment was simulated and found to be \SI{3.7}{\percent} \cite{VicenteBarretoPinto:2267848}. Hence, a value of \SI{4}{\percent} was used in this study and is added throughout this paper. The cumulative effect of the cross-coupling addition can be seen in figure \ref{fig: pixelToTResponse}, where a comparison of the pixel ToT as a function of track distance is shown for both simulation and data. Again, the data shown is for clusters in the column direction of the CLICpix and for perpendicular track incidence, where the track passes close to the centre of the pixel, in both x and y) in order to match the 2D TCAD simulation. There is a general agreement between simulation and data, despite several factors which are not taken into account in the simulations, particularly Landau fluctuations of the deposited charge as well as noise.

\subsection{Bias scan}

\begin{figure}[b!]
	\centering 
	\includegraphics[width=.45\textwidth]{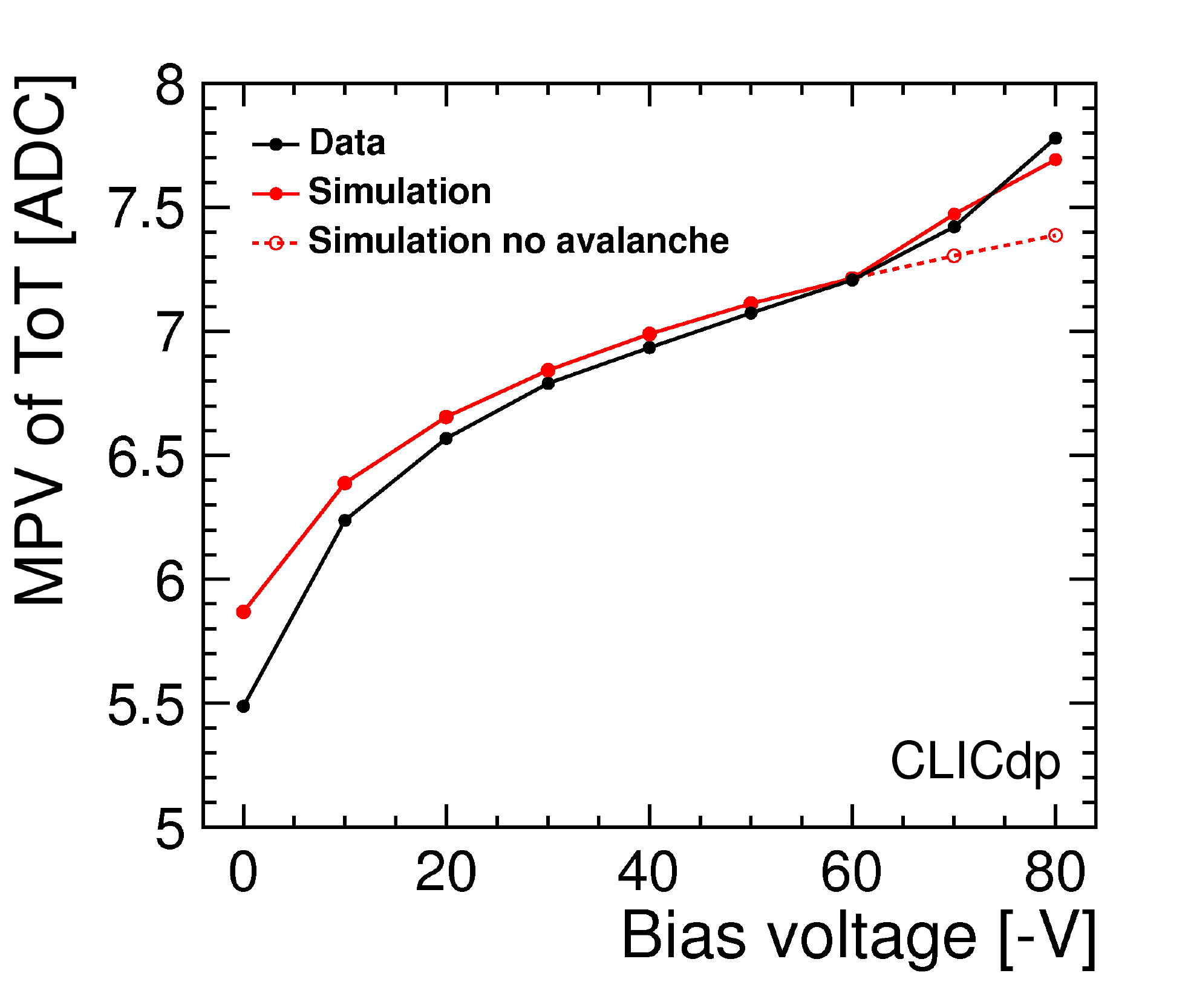}
	\caption{Most probable ToT value for single pixel clusters as a function of bias, for data (black) and simulation (red). The dashed red line represents the simulations without avalanche model.}
	\label{fig: bias comp}
\end{figure}

An advantage of using TCAD simulation is the ability to describe the detector response over a wide range of bias voltages and field configurations. The results for different bias voltages are shown in figure \ref{fig: bias comp}, where the most probable value of the ToT distribution for single pixel clusters is compared between simulation and data. While both exhibit the typical increase in collected charge expected from the extension of the depletion region between 0 and -60~V, the data beyond this range shows a marked rise in the amount of charge measured by the sensor. This rise is only reproduced in simulation with the addition of an avalanche model to TCAD, which reproduces the observed behaviour. The avalanche model used was the University of Bologna impact ionization model for silicon \cite{1419171}. High field regions in the vicinity of the deep n-well are thus producing avalanche conditions for the multiplication of charge close to the implants. The systematic deviation of the simulations above the data for low bias voltages could possibly arise from the discrepancies between the avalanche model used and a real sensor; the simulation overestimates the amount of gain at lower voltages.

\subsection{Charge collection for angled tracks}

\begin{figure}[t!]
	\centering
	\begin{tabular}{@{}p{0.5\linewidth}@{\quad}p{0.5\linewidth}@{}}
		\subfigimg[width=.45\textwidth]{(a)}{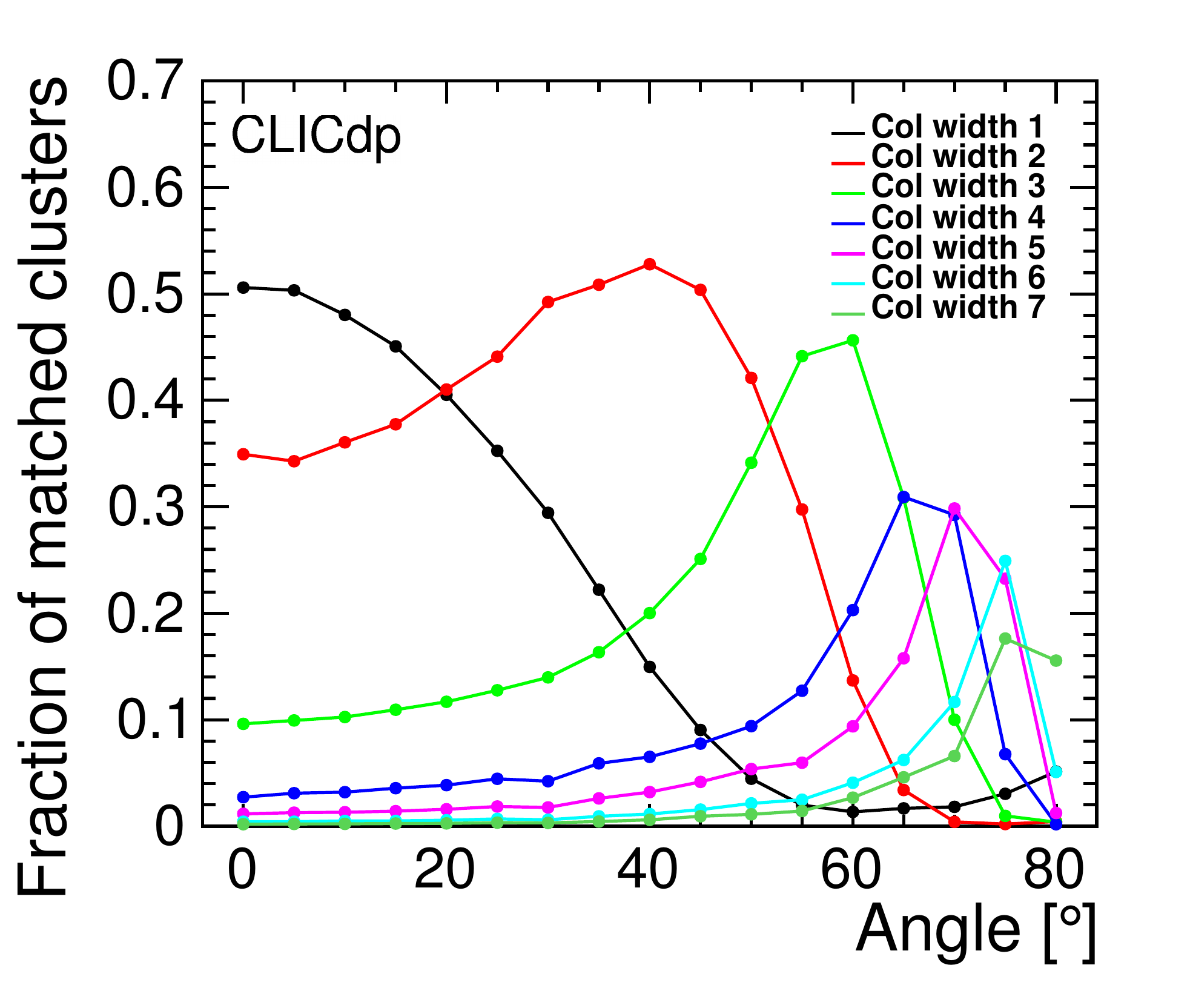} &
		\subfigimg[width=.45\textwidth]{(b)}{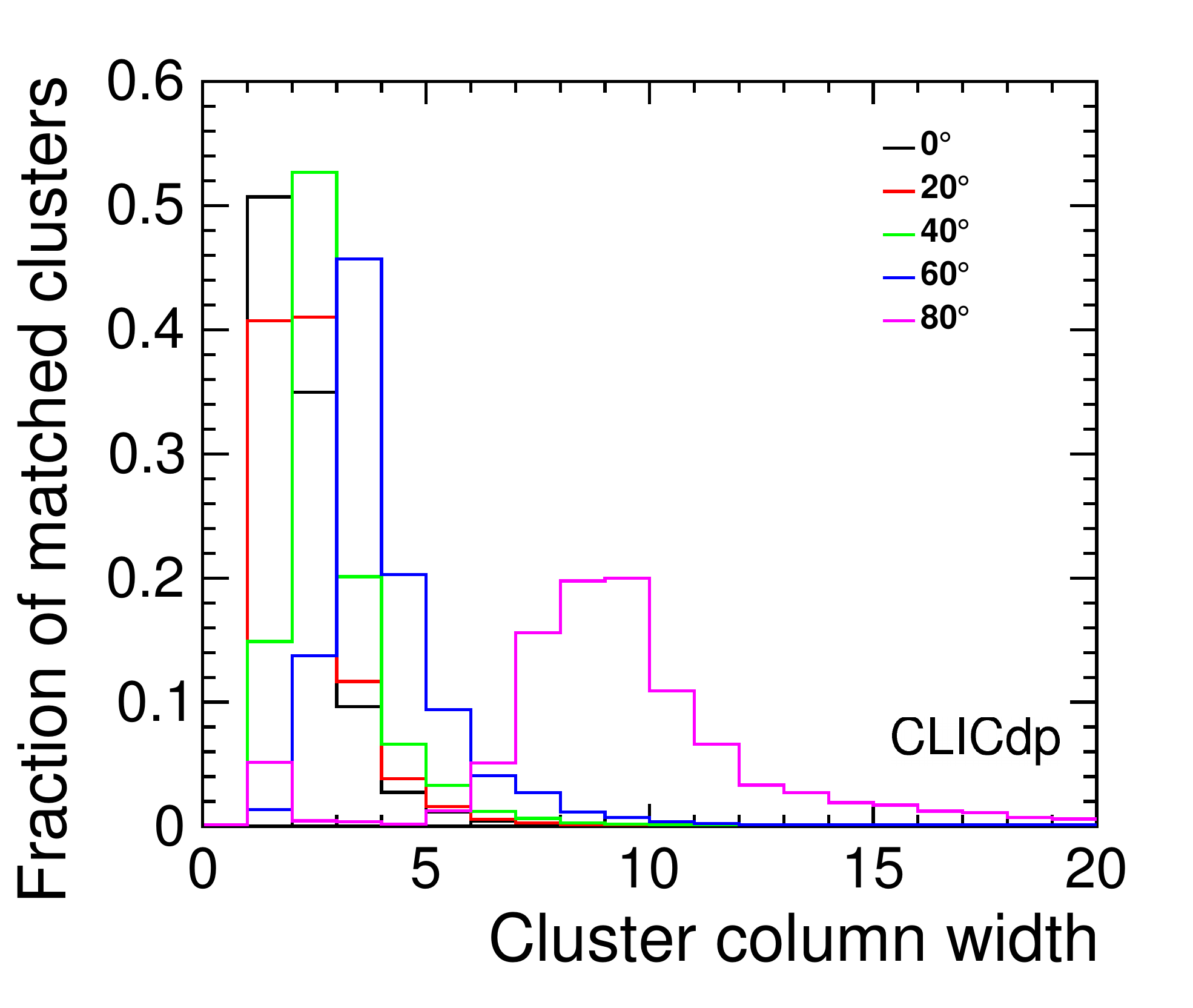}
	\end{tabular}
	\caption{(a) Cluster width distributions in the direction of rotation, for various angles. (b) Fraction of n-pixel column widths as a function of angle.}
	\label{fig: ang variation}
\end{figure}

Results have so far been presented for tracks at perpendicular incidence to the detector surface. For the proposed CLIC vertex detector however, track angles of up to \ang{75} are expected \cite{AlipourTehrani:2254048}. Thus, response of capacitively coupled assemblies in this angular range are studied to better understand their performance.

\begin{figure}[b!]
	\centering
	\begin{tabular}{@{}p{0.5\linewidth}@{\quad}p{0.5\linewidth}@{}}
		\subfigimg[width=.45\textwidth]{(a)}{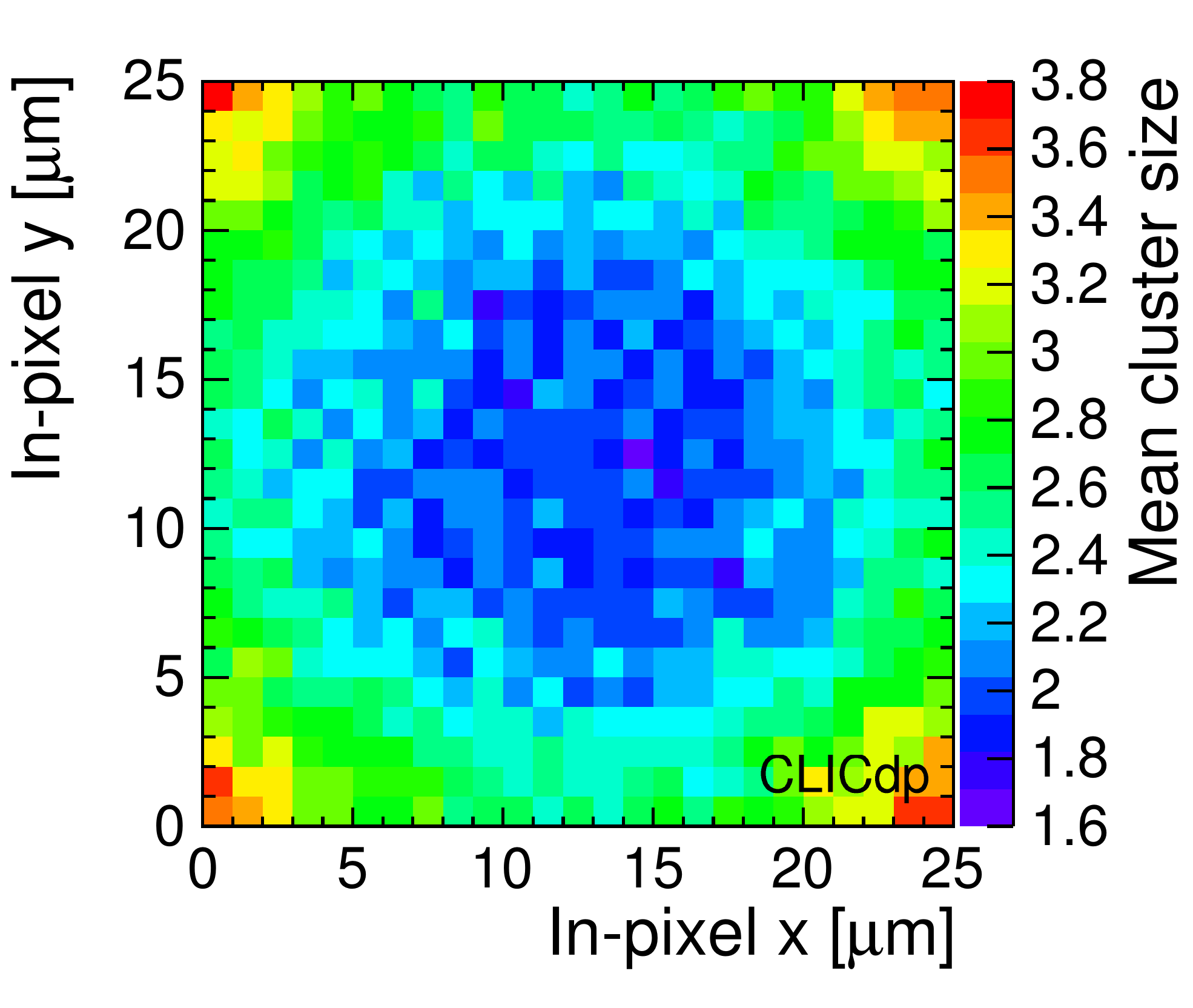} &
		\subfigimg[width=.45\textwidth]{(b)}{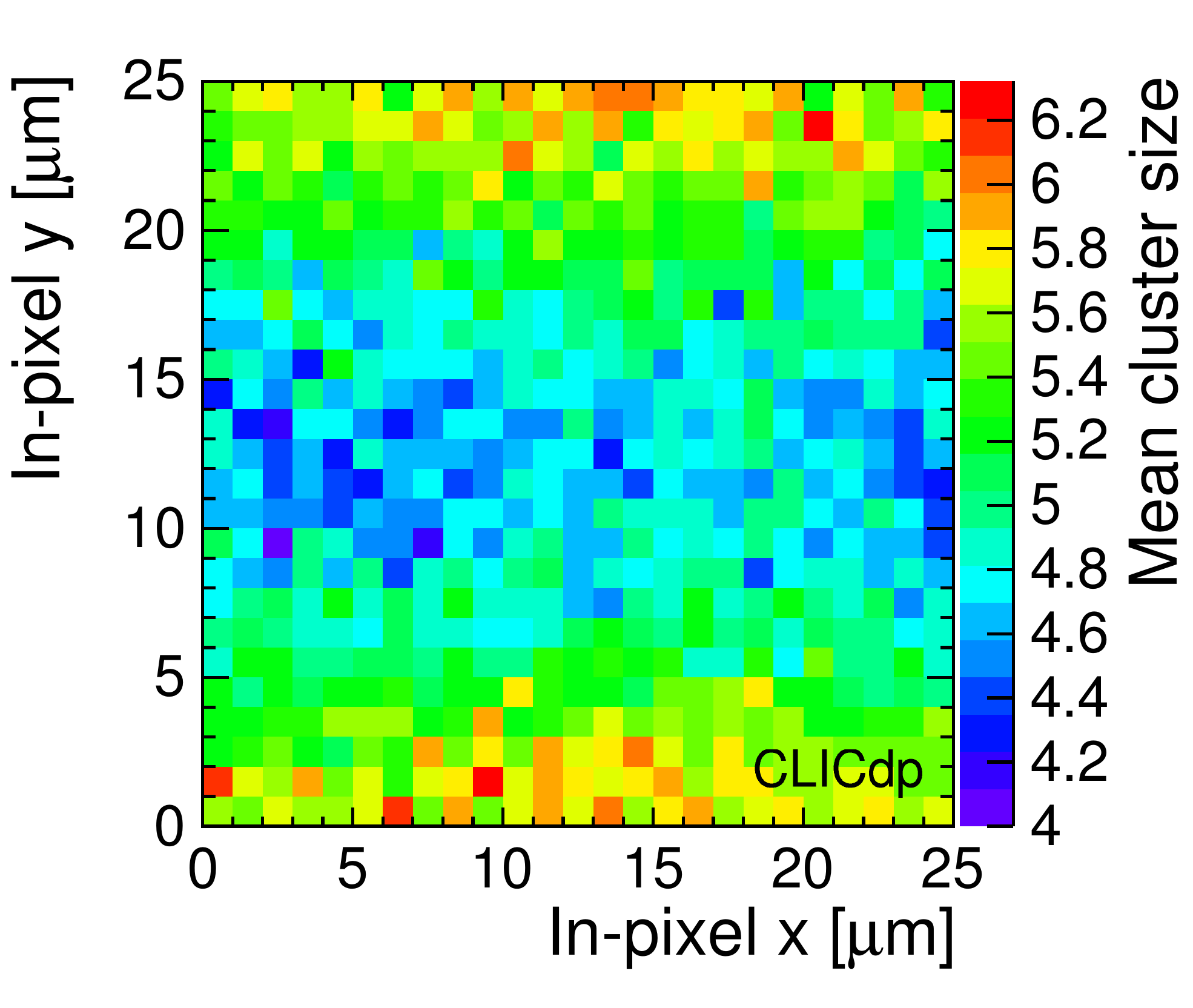}
	\end{tabular}
	\caption{Mean cluster size as a function of track position within the pixel cell for (a) \ang{0} and (b) \ang{60}.}
	\label{fig: inPix mean clust size}
\end{figure}

The cluster width distributions (perpendicular to the rotation axis, x) for an array of angles are shown in figure \ref{fig: ang variation}, along with the fraction of n-pixel clusters as a function of angle. Given the limited depth of the depleted region, even at angles of up to \ang{60} the majority of clusters display widths of less than 5, while those produced by tracks almost parallel to the detector surface have a large variance in their observed size. Looking in more detail at the cluster size as a function of track position within the pixel cell, the charge sharing behaviour of the sensor can be observed. Figure~\ref{fig: inPix mean clust size} shows the in-pixel mean cluster size for data at \ang{0} and \ang{60}. At perpendicular incidence the mean cluster size in the centre of the pixel takes values of around 2. This can be attributed to the small pixel size generating many multi-pixel clusters. For a track angle of \ang{60}, the behaviour is as expected from the geometry of the setup, showing stripes of certain cluster sizes perpendicular to the rotation axis.

\begin{figure}[t!]
	\centering
	\begin{tabular}{@{}p{0.5\linewidth}@{\quad}p{0.5\linewidth}@{}}
		\subfigimg[width=.45\textwidth]{(a)}{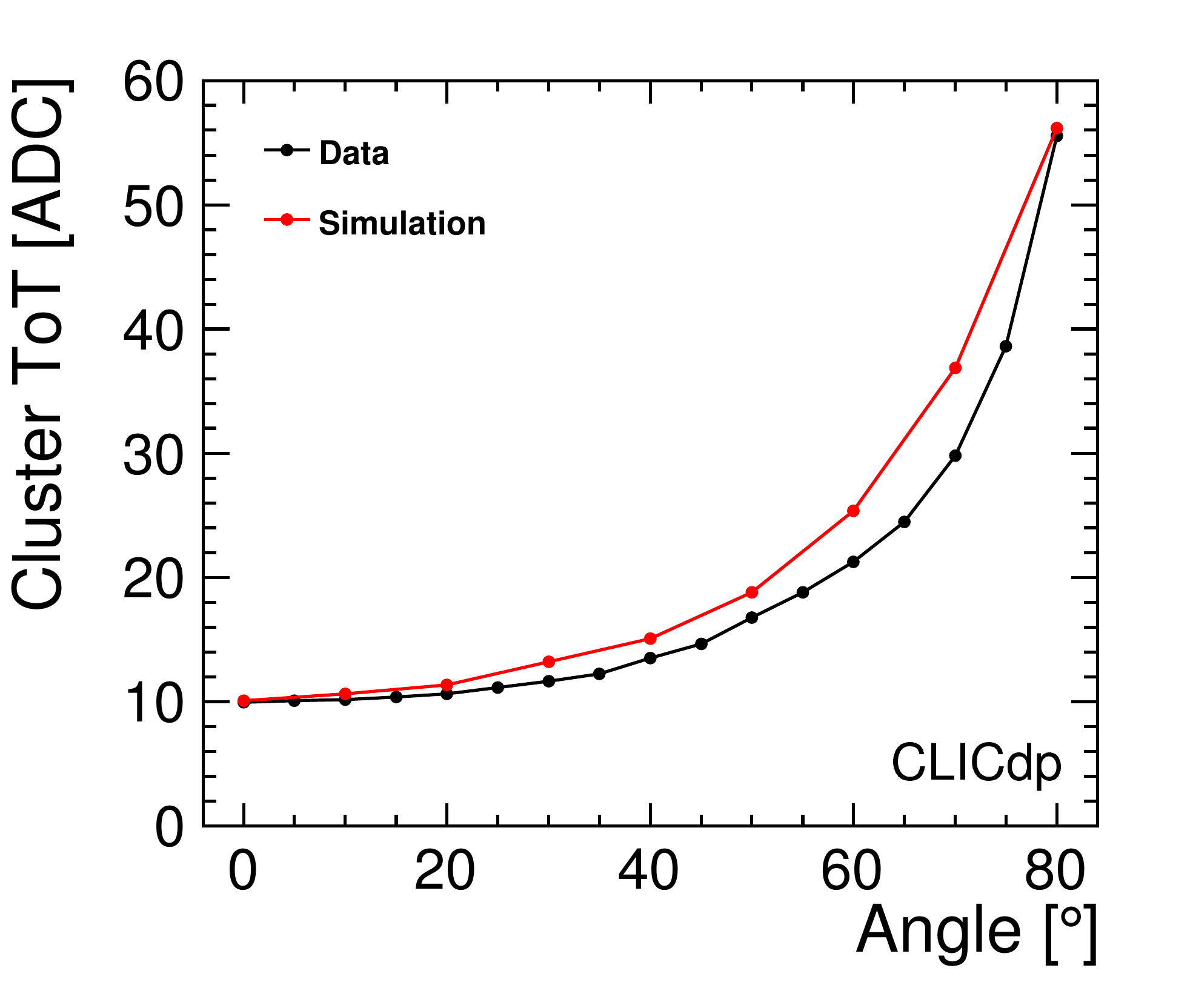} &
		\subfigimg[width=.45\textwidth]{(b)}{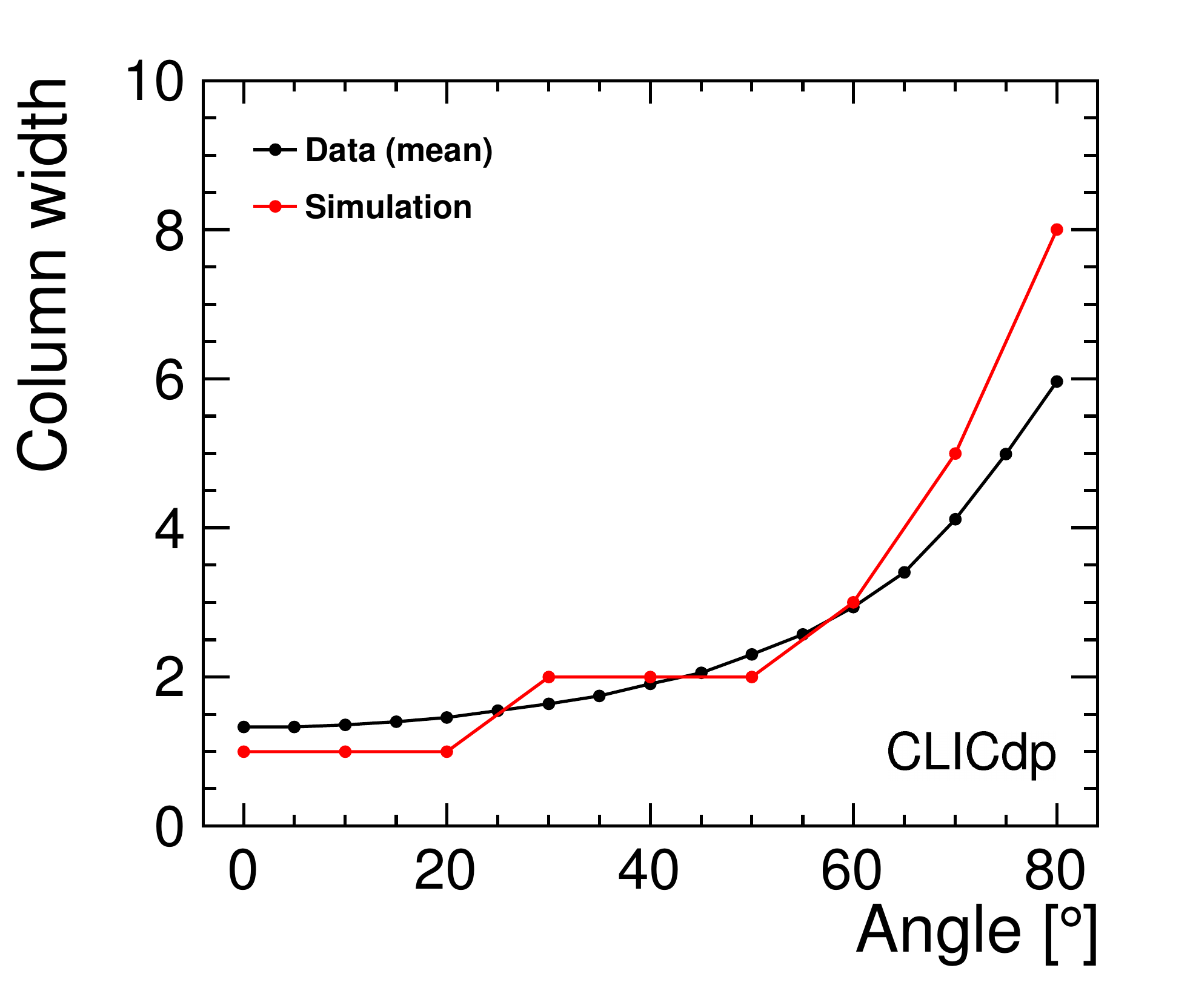}
	\end{tabular}
	\caption{(a) Cluster ToT distribution as a function of angle. (b) Column width as a function of angle.}
	\label{fig: MPV ToT angle}
\end{figure}

\pagebreak

The cluster ToT and the column width over the angular range, with and without the effect of cross-coupling (CC) in the simulation, are shown in figure \ref{fig: MPV ToT angle}. The cluster ToT for the data is the mean from a Gaussian fit because the distributions do not look like a Landau distribution due to the limited resolution of the ToT counter. For the simulation, the energy deposition is set to the 80 electron-hole pairs per \si{\um} using a constant charge deposition as a function of depth and there is only one simulation per data point. In the simulations a threshold, to be part of the cluster, is defined as the point at which the fit in figure \ref{fig: calibration} crosses the x-axis. The simulations show a general agreement with the beam test data even at the higher angles. The exception is the column width at \ang{80} where the simulation value is much higher. This is due to the cluster finding and matching used for the test beam results not being optimised for high incidence angles. In both plots of figure \ref{fig: MPV ToT angle}, the data is restricted to clusters with a row width of one and an additional cut on track positions within $\pm$\SI{2.5}{\micro\m} of the pixel centre was made to match the 2D simulations. For the cluster ToT this was applied in both column and row directions, while for the column width an additional cut was only applied to the row-direction, creating a strip in the column-direction.

\subsection{Active depth}

\begin{figure}[t!]
	\centering 
	\includegraphics[width=.45\textwidth]{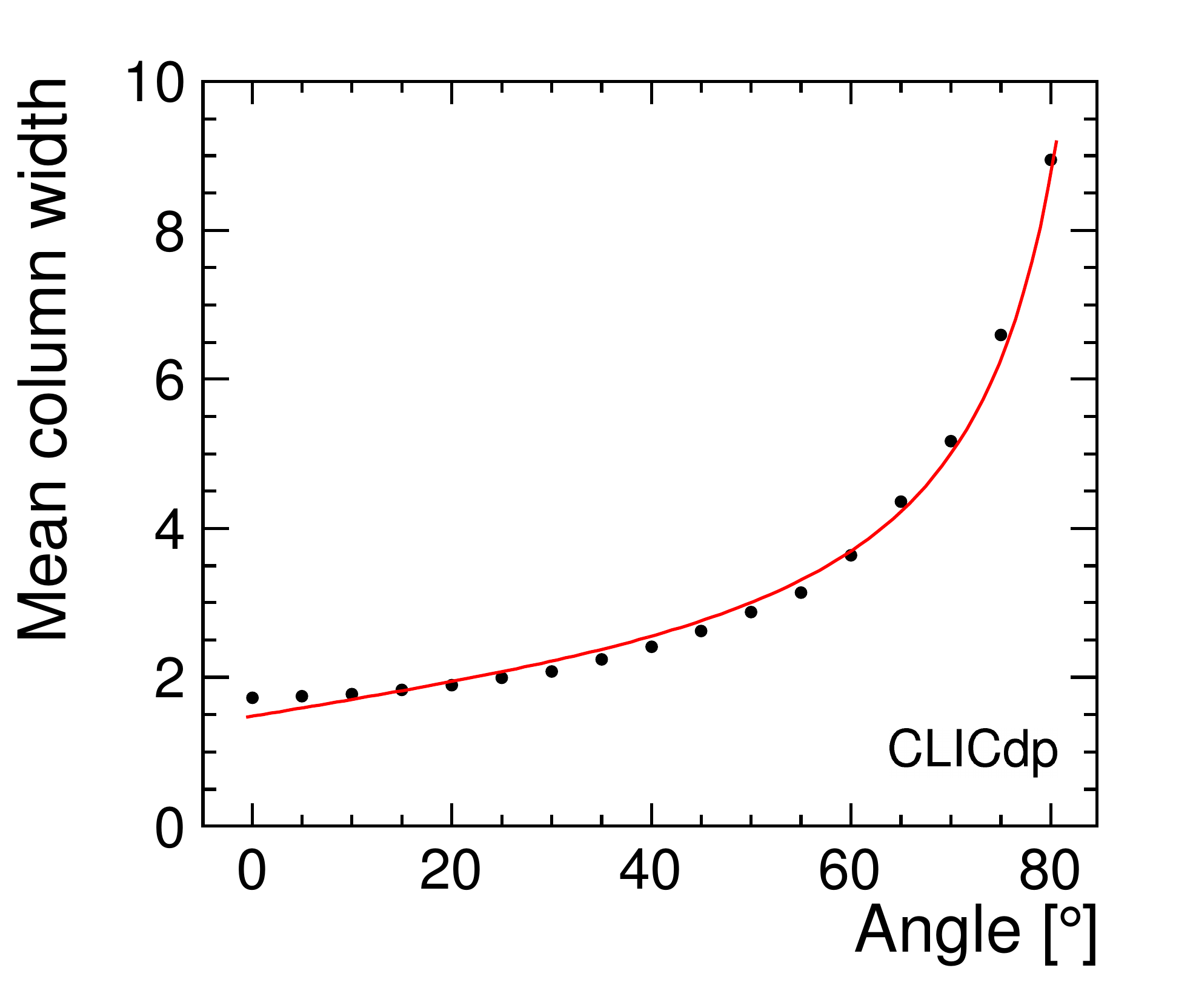}
	\caption{The measured mean column width as a function of rotation angle with the fit shown in red.}
	\label{fig: mean col width}
\end{figure}

While the simulation studies in \ref{sec:CCModes} have shown both the timing characteristics and the contribution of charge collected from the undepleted bulk, the active depth of the sensor will be sensitive to additional factors such as the integration time of the electronics and the operating threshold of the readout ASIC. In order to get an estimate of the effective active depth, a geometric approximation is taken which makes use of the known parameters including the pitch, $p$, and the rotation angle, $\theta$. These quantities can be related to the cluster width in the direction of rotation by the following equation: 

\begin{equation}
	\label{eq: active depth}
	\text{column width} = \tan{(\theta)} \frac{d}{p} + c,
\end{equation}

\noindent where $c$ is the cluster width at perpendicular incidence, and $d$ is the active depth. This is used to fit the data in figure \ref{fig: mean col width}, which shows the mean cluster width as a function of angle. Unlike in figure \ref{fig: MPV ToT angle} (b) the data has only been restriction applied if for clusters of row width of one because there is no comparison with simulation. From this, a value of $\approx$ \SI{30}{\micro\m} was found for the active depth - almost 3 times as large as the simulated depletion depth of \SI{11.5}{\micro\m} in section \ref{sec:Efield} - indicating a significant contribution from diffusion to the charge collection, in agreement with results in section \ref{sec:CCModes}. The result is also compatible with the TPA-eTCT measurements of the active depth of \SI{25}{\micro\m} \cite{GARCIA201769}, considering the different integration times of the CLICpix ($\sim$\SI{30}{\nano\s}) and the TPA-eTCT measurements (\SI{20}{\nano\s}).

\subsection{Single hit resolution}

\begin{figure}[b!]
	\centering 
	\includegraphics[width=.45\textwidth]{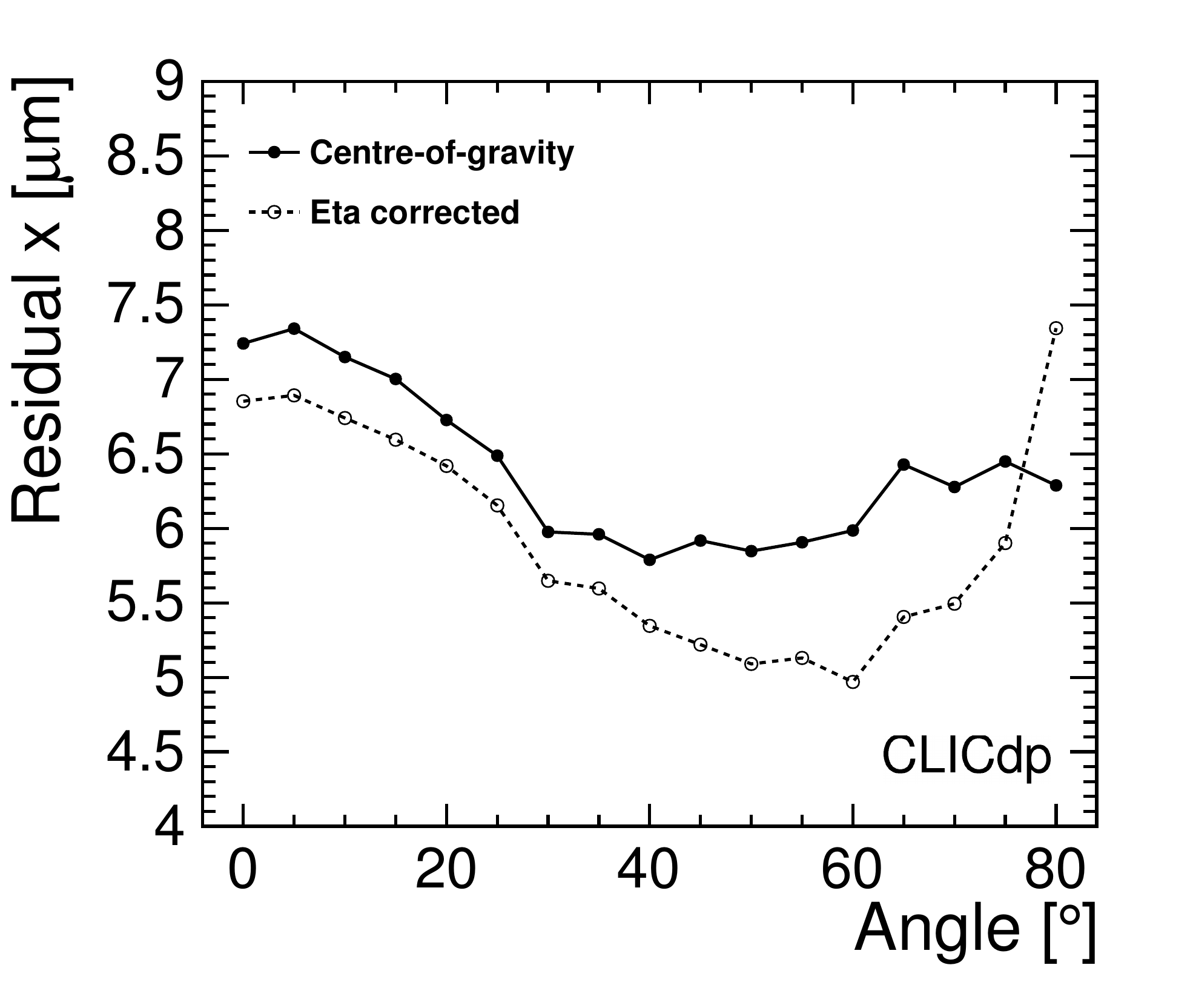}
	\caption{Single hit residual, from a Gaussian fit, in the x-direction at different angles, for centre-of-gravity (bold) and eta-corrected (dashed) cluster positions.}
	\label{fig: resolution}
\end{figure}

One of the challenging requirements for the CLIC vertex detector is to provide a single hit resolution of \SI{3}{\micro\m}, while retaining a time-tagging precision of $<$ \SI{10}{\nano\s}. The single hit resolution of the capacitively coupled CCPDv3-CLICpix assembly has been measured over the full angular range, as shown in figure \ref{fig: resolution}. The resolution of the DUT is defined here as the standard deviation of a Gaussian fitted to the residual distribution for all clusters within \SI{99.7}{\percent} of the statistics ($\pm3\sigma$ of the mean), and can be seen versus angle, both before and after eta corrections, in figure \ref{fig: resolution}. The eta corrections are applied separately for each incidence angle.  From \ang{0} onwards the resolution value decreases and plateaus at around \ang{50} because the cluster size is optimal for the centre-of-gravity algorithm. After \ang{60} the cluster sizes become much larger and a new cluster position algorithm, such as the head-to-tail algorithm, should be used. The reason for the eta correction making the resolution worse than the non-corrected value at \ang{80} is that the track impact position resolution is worse at higher angles ($\sigma_{track}=$\,\SI{1.7}{\micro\m} at \ang{20} and $\sigma_{track}=$\,\SI{9.2}{\micro\m} at \ang{80}). This increases the potential for mismatch between the centre-of-gravity position and the track position impacting on the eta-correction. The RMS of the residual distributions covering \SI{99.7}{\percent} of the data is \SIrange{8}{9}{\micro\m} for all angles.

\subsection{Single hit efficiency}

\begin{figure}[t!]
	\centering 
	\includegraphics[width=.45\textwidth]{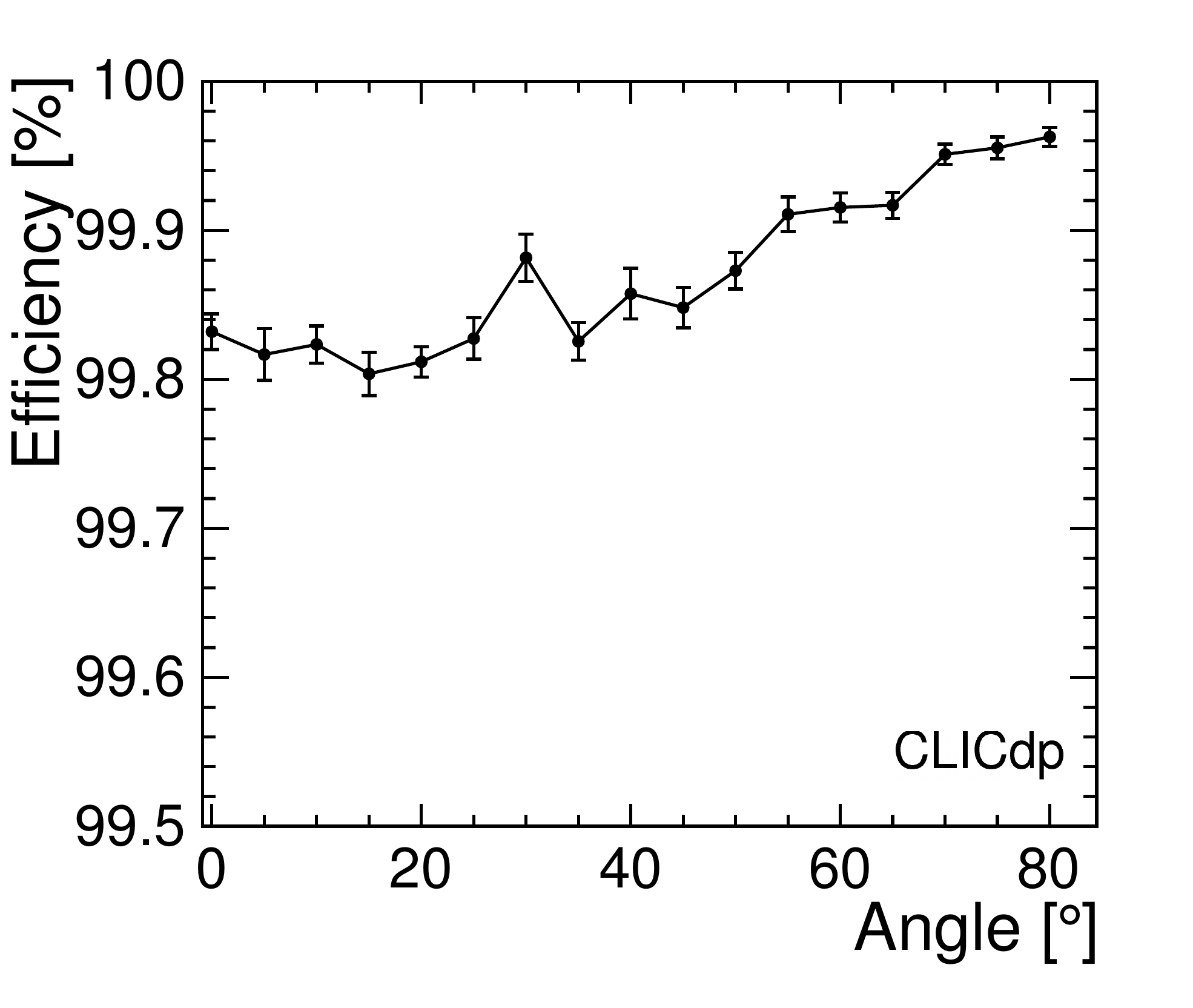}
	\caption{Measured single hit efficiency of the DUT as a function of angle with the error bars denoting the statistical uncertainty.}
	\label{fig: efficiency}
\end{figure}

Of importance for all particle physics experiments is the detector efficiency, which should generally be greater than \SI{99}{\percent} in each sensor plane to help achieve precision vertex tagging. The efficiency for such assemblies has already been measured previously, \cite{Tehrani:2048684}, but is presented here over the full angular range. The efficiency is calculated using the number of associated tracks divided by the total number of tracks that pass through the DUT. An associated track is one that has passed the several cuts previously mentioned in Section \ref{sec: TrackPerform}. Figure \ref{fig: efficiency} shows that the device tested had efficiencies above \SI{99.8}{\percent} over the full acceptance, with a small improvement at higher angles where several pixels are crossed.

\section{Prospects for improved performance}

\begin{figure}[t!]
	\centering
	\begin{tabular}{@{}p{0.5\linewidth}@{\quad}p{0.5\linewidth}@{}}
		\subfigimg[width=.45\textwidth]{(a)}{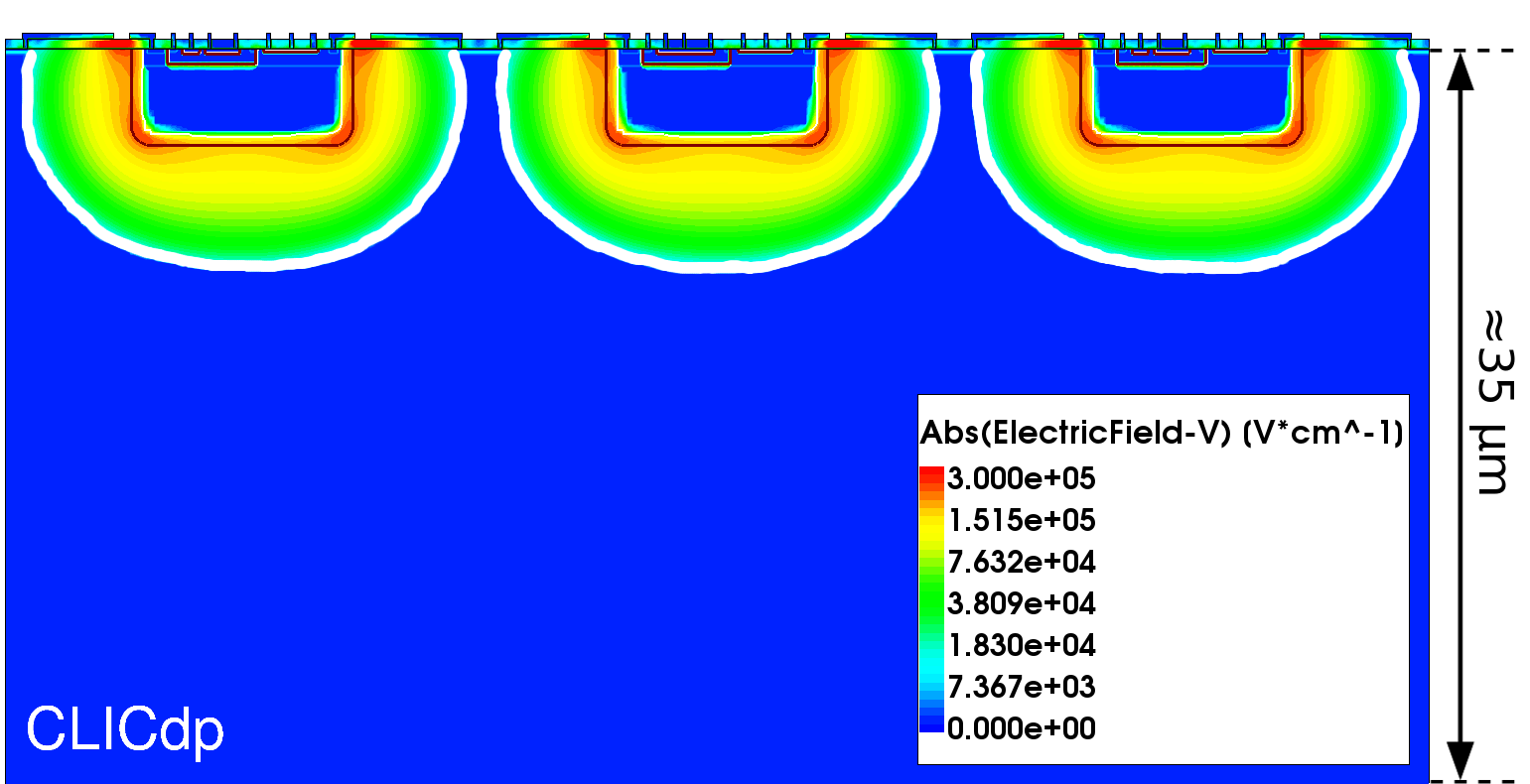} &
		\subfigimg[width=.45\textwidth]{(b)}{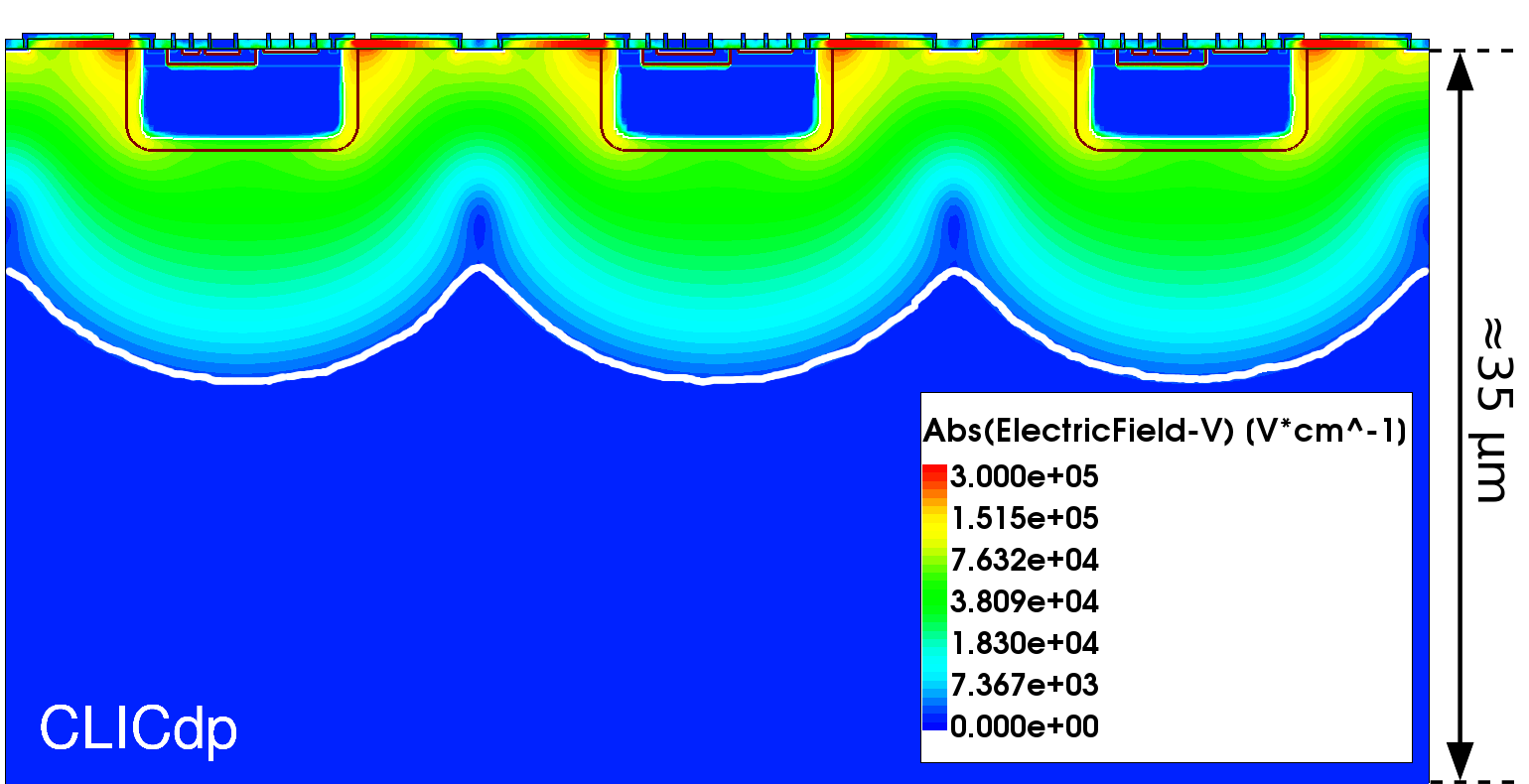} \\
		\subfigimg[width=.45\textwidth]{(c)}{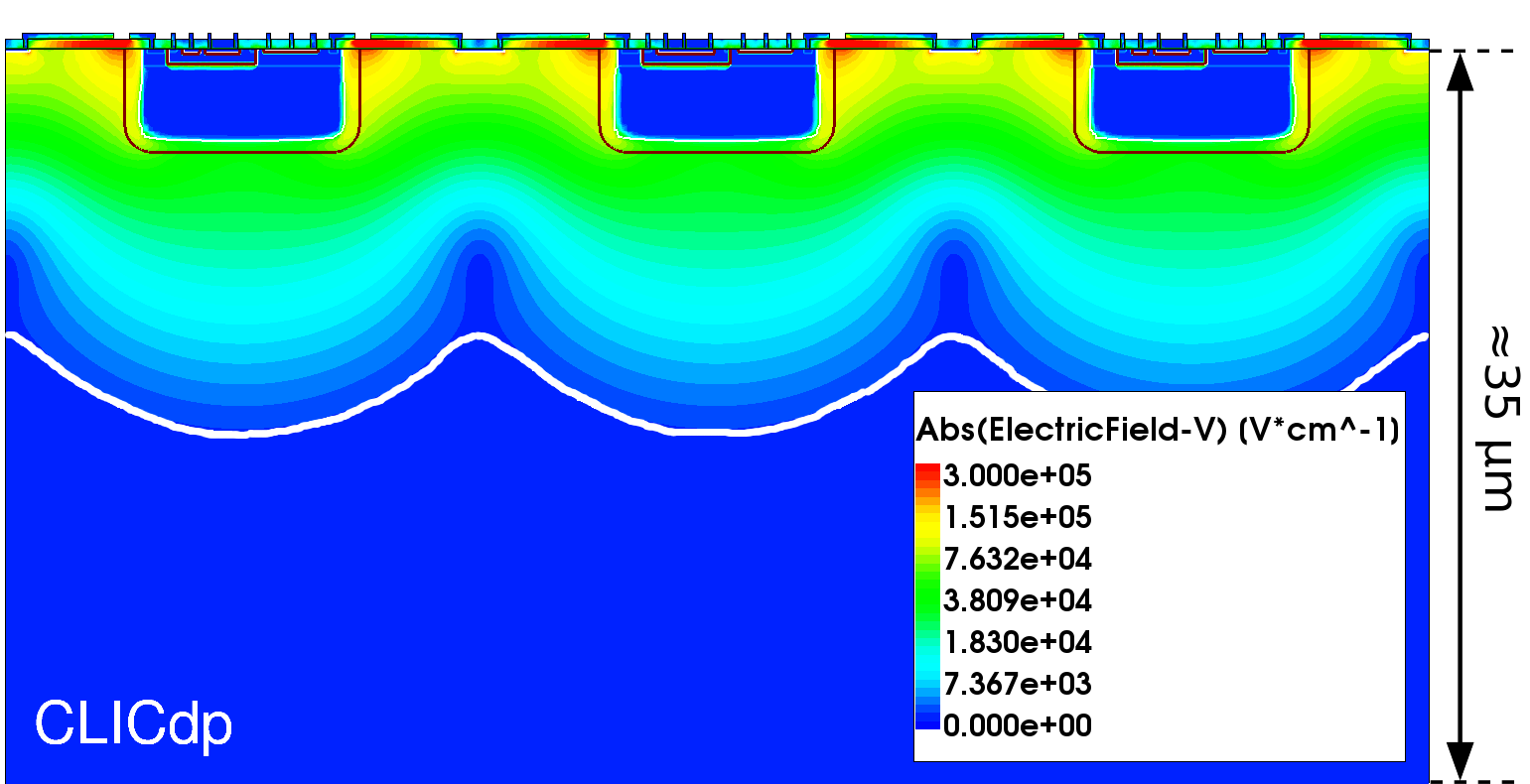} &
		\subfigimg[width=.45\textwidth]{(d)}{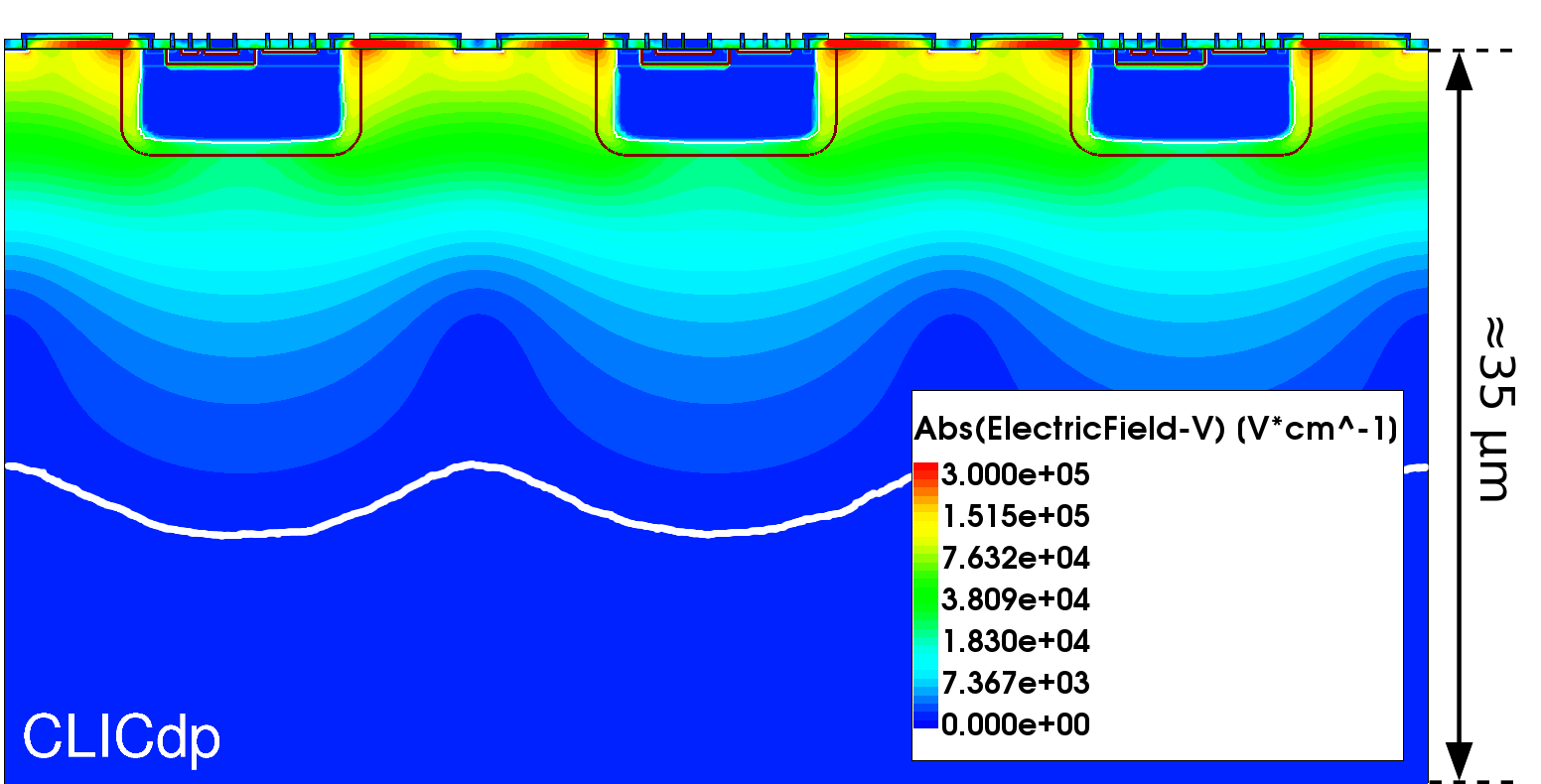}
	\end{tabular}
	\caption{Field maps showing the absolute value of the electric field, all to the same scale, at \SI{-60}{\volt} for different bulk resistivity values: (a) \SI{10}{\ohm\cm}, (b) \SI{80}{\ohm\cm}, (c) \SI{200}{\ohm\cm} and (d) \SI{1000}{\ohm\cm}. The white line indicates the border of the depletion region.}
	\label{fig: Efield res comp}
\end{figure}

Given the general agreement between TCAD simulations and the performance of HV-CMOS detectors shown above, the simulations can further be used to inform future sensor designs, with several prospects for improved performance. One such area where this can be of aid is in extrapolating the current results to silicon substrates of higher resistivity, where it can be expected that, given a larger depletion region, charge collected by drift will contribute more to the total charge collection. To quantify this expectation, a series of simulations was carried out for several bulk resistivity values: \SI{10}{\ohm\cm} (nominal value), \SI{80}{\ohm\cm}, \SI{200}{\ohm\cm} and \SI{1000}{\ohm\cm} with a substrate depth of \SI{250}{\micro\m}. From figure \ref{fig: Efield res comp} the depletion region (indicated by the white line) and the absolute value of the electric field within the sensors can be seen, showing an extension of the field further into the bulk for higher resistivities. In addition, the high field regions around the deep n-well (red and yellow) recede, spreading instead more laterally and indicating a faster charge collection when the MIP passes between two pixels.

\begin{figure}[t!]
	\centering
	\begin{tabular}{@{}p{0.5\linewidth}@{\quad}p{0.5\linewidth}@{}}
		\subfigimg[width=.45\textwidth]{(a)}{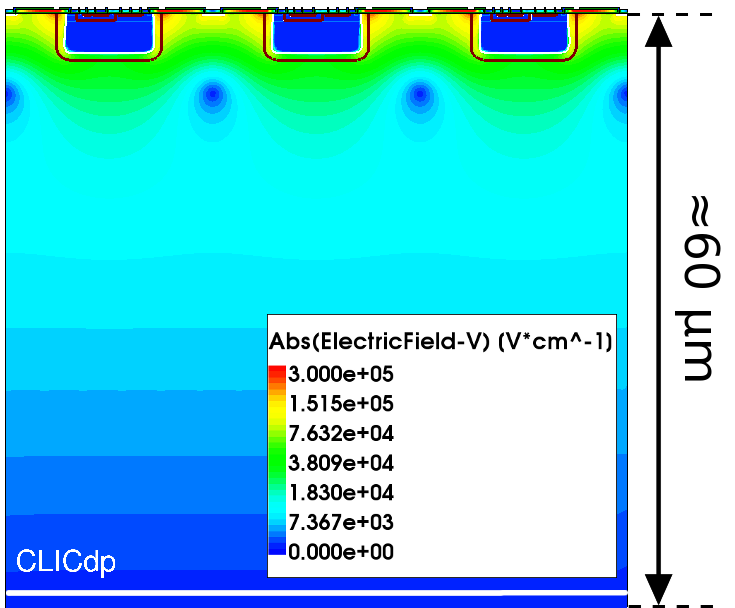} &
		\subfigimg[width=.45\textwidth]{(b)}{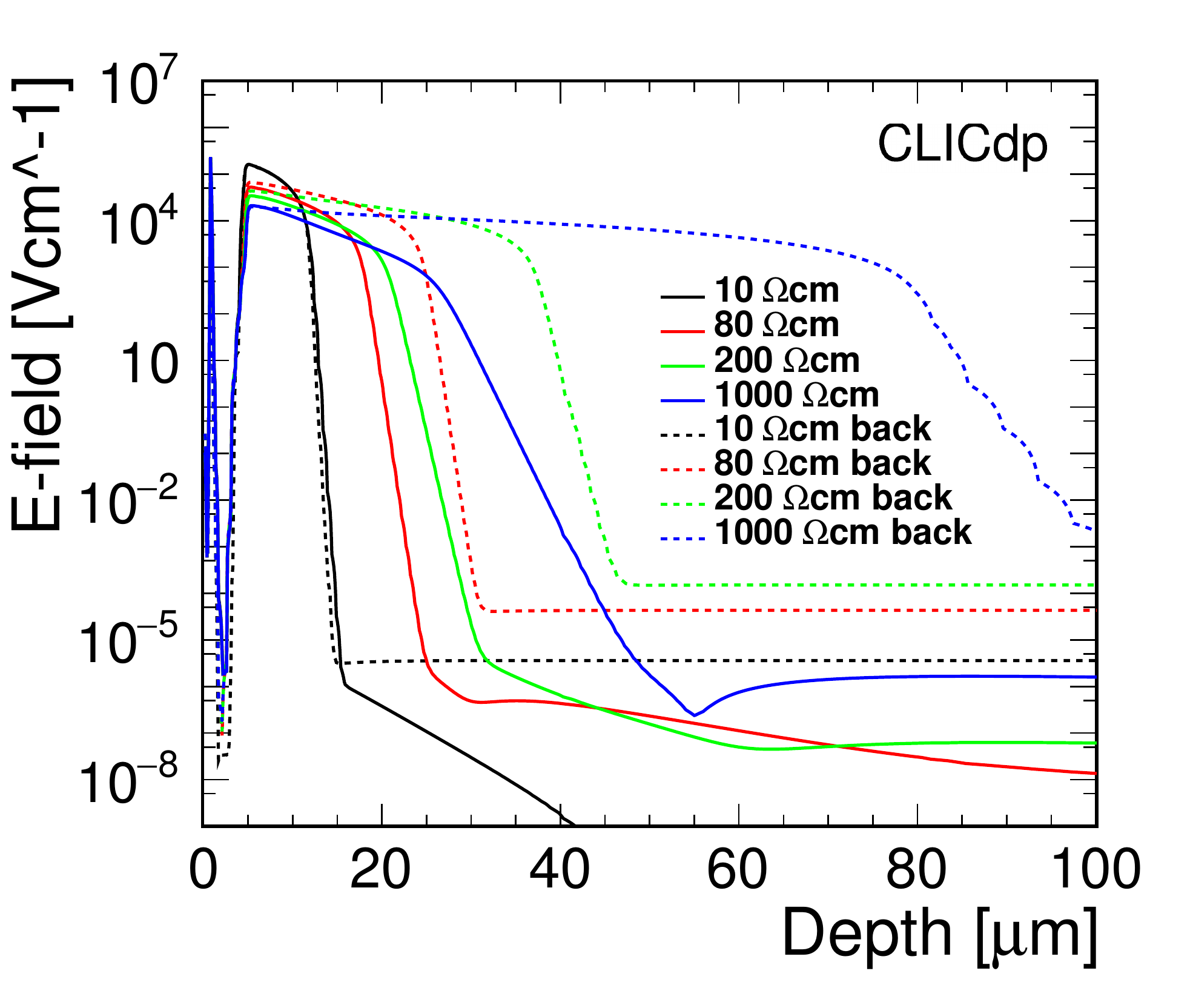}
	\end{tabular}
	\caption{(a) Field map of the absolute value of the electric field at \SI{-60}{\volt}, for a back-biased scheme with bulk resistivity \SI{1000}{\ohm\cm}. (b) Absolute value of the electric field along a line through the pixel centre as a function of depth for various bulk resistivities, for top (bold) and back (dashed) biasing schemes.}
	\label{fig: Efield res comp back}
\end{figure}

In addition to higher resistivities, a further way to improve the performance of such HV-CMOS sensors would be to bias the device from the backside instead of the topside. This causes the depletion region and electric field to extend further into the bulk (figure \ref{fig: Efield res comp back} (a)), increasing the speed and quantity of charge collected. By taking a profile through the centre of the simulated structures, a comparison of the electric field as a function of depth can be made, figure \ref{fig: Efield res comp back} (b), showing that the higher the resistivity the greater the difference between topside and backside biasing schemes. At a bulk resistivity of \SI{10}{\ohm\cm} very little difference in electric field depth is observed; only at high values of bulk resistivity do the effects become more prominent.

\pagebreak

\begin{figure}[!]
	\centering 
	\begin{tabular}{@{}p{0.5\linewidth}@{\quad}p{0.5\linewidth}@{}}	
		\subfigimg[width=.45\textwidth]{(a)}{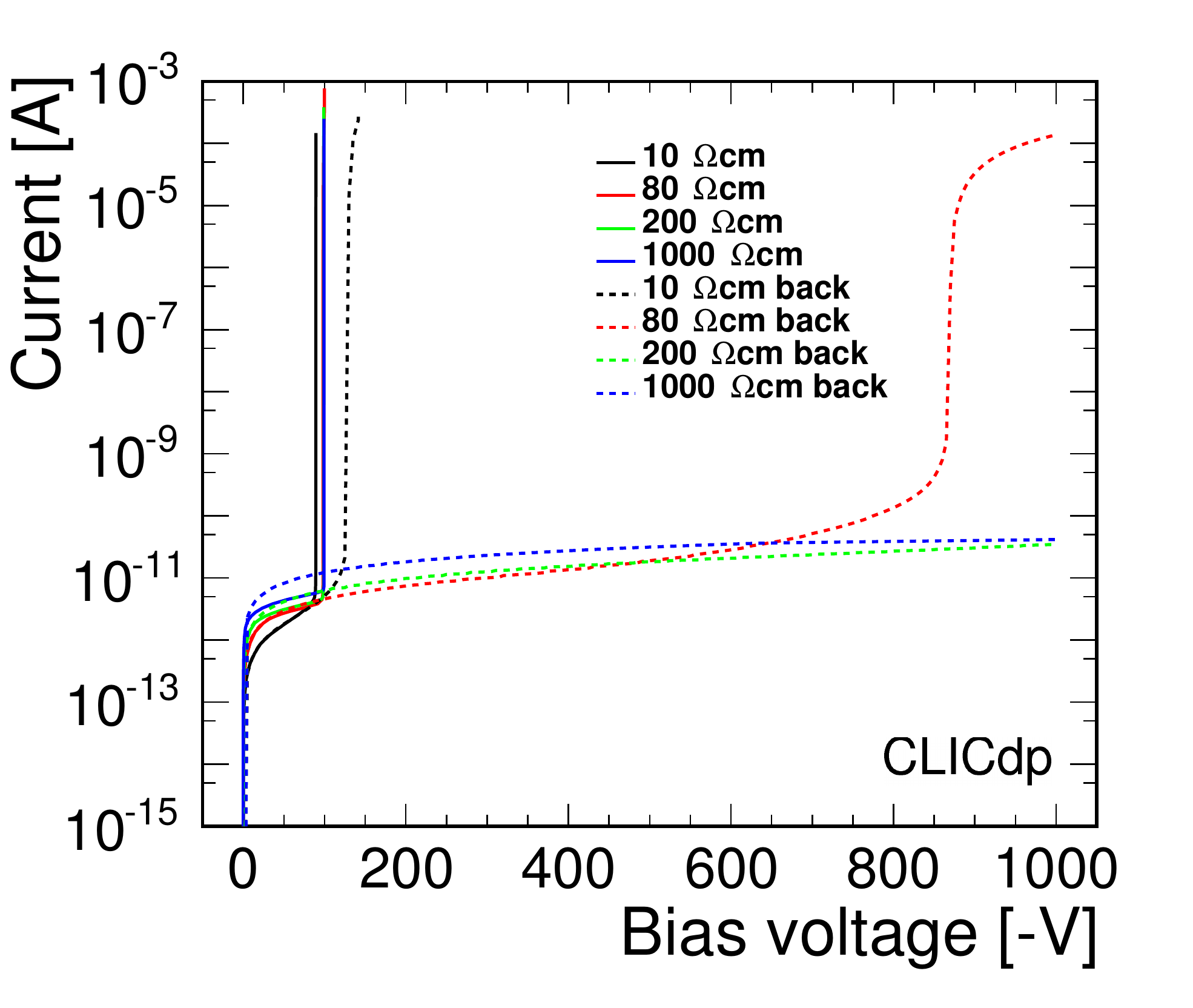}
	\end{tabular}	
	\caption{Current as a function of bias showing the breakdown for different bulk resistivities and the top and back biasing schemes.}
	\label{fig: IV res comp}
\end{figure}

\begin{figure}[!]
	\centering 
	\begin{tabular}{@{}p{0.50\linewidth}@{\quad}p{0.5\linewidth}@{\quad}p{0.5\linewidth}@{}}	
		\subfigimg[width=.43\textwidth]{(a)}{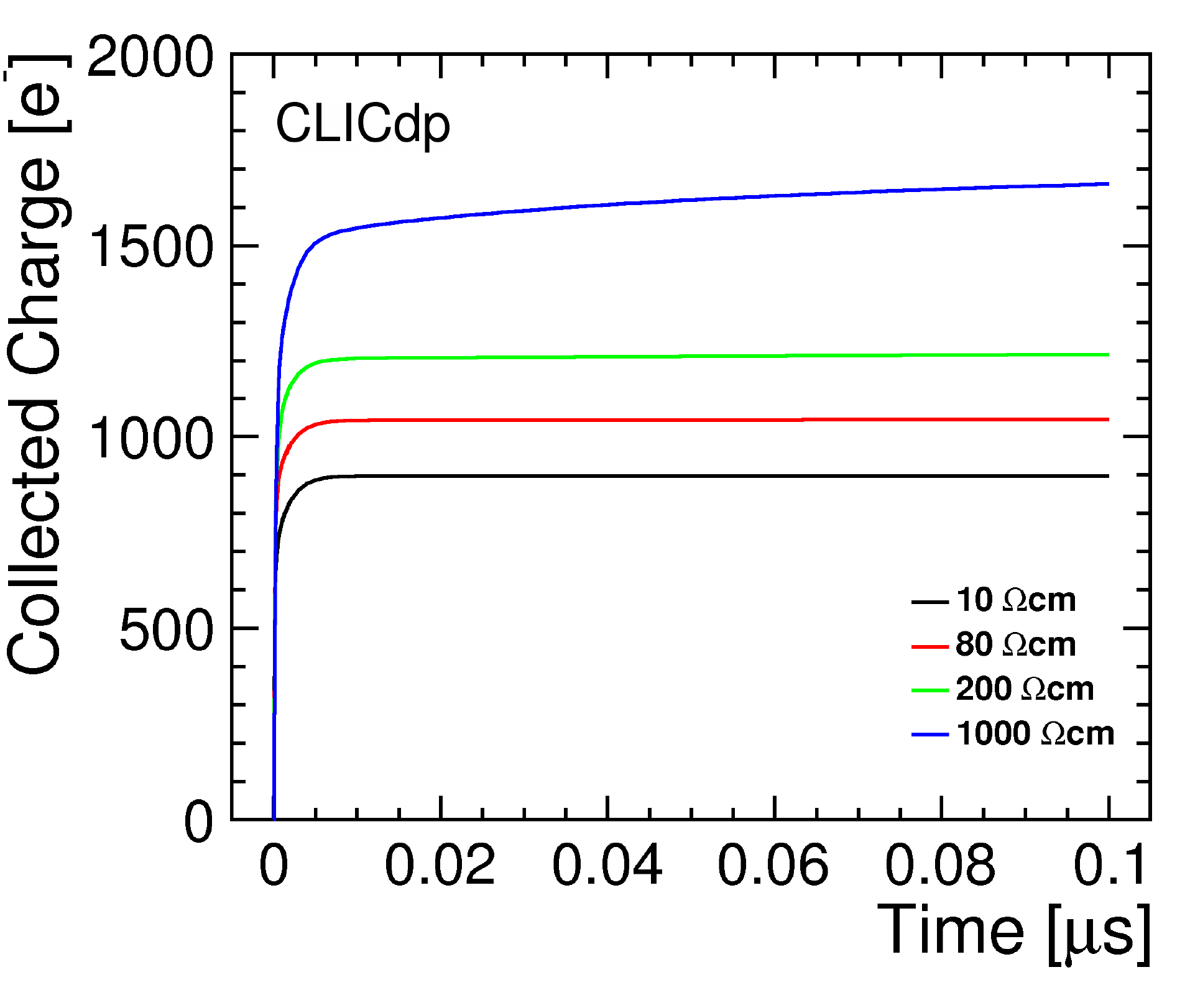} &
		\subfigimg[width=.43\textwidth]{(b)}{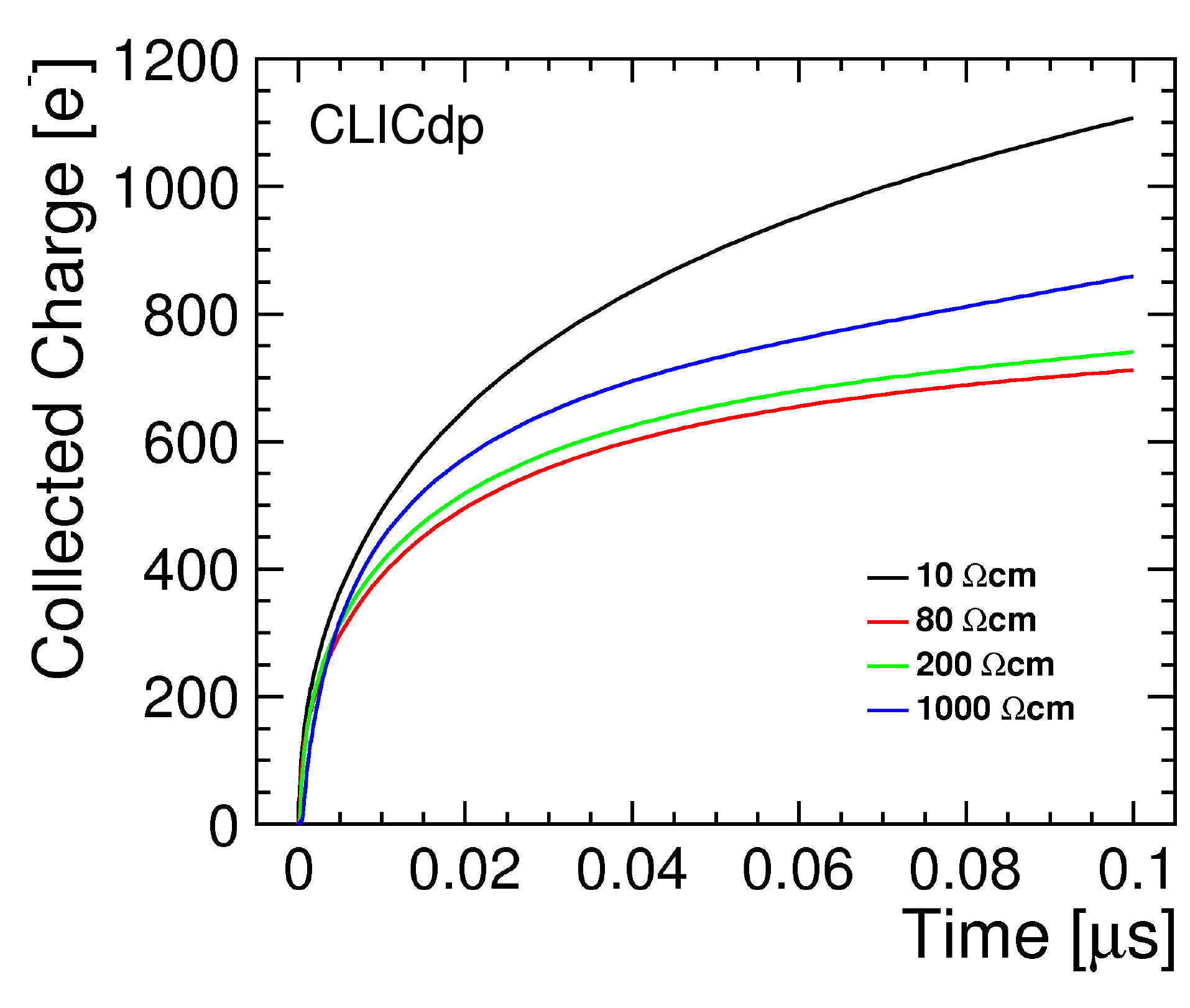} \\
		% can't get this image to centre so a fudge will do for now
		\hspace{4cm}\subfigimg[width=.43\textwidth]{(c)}{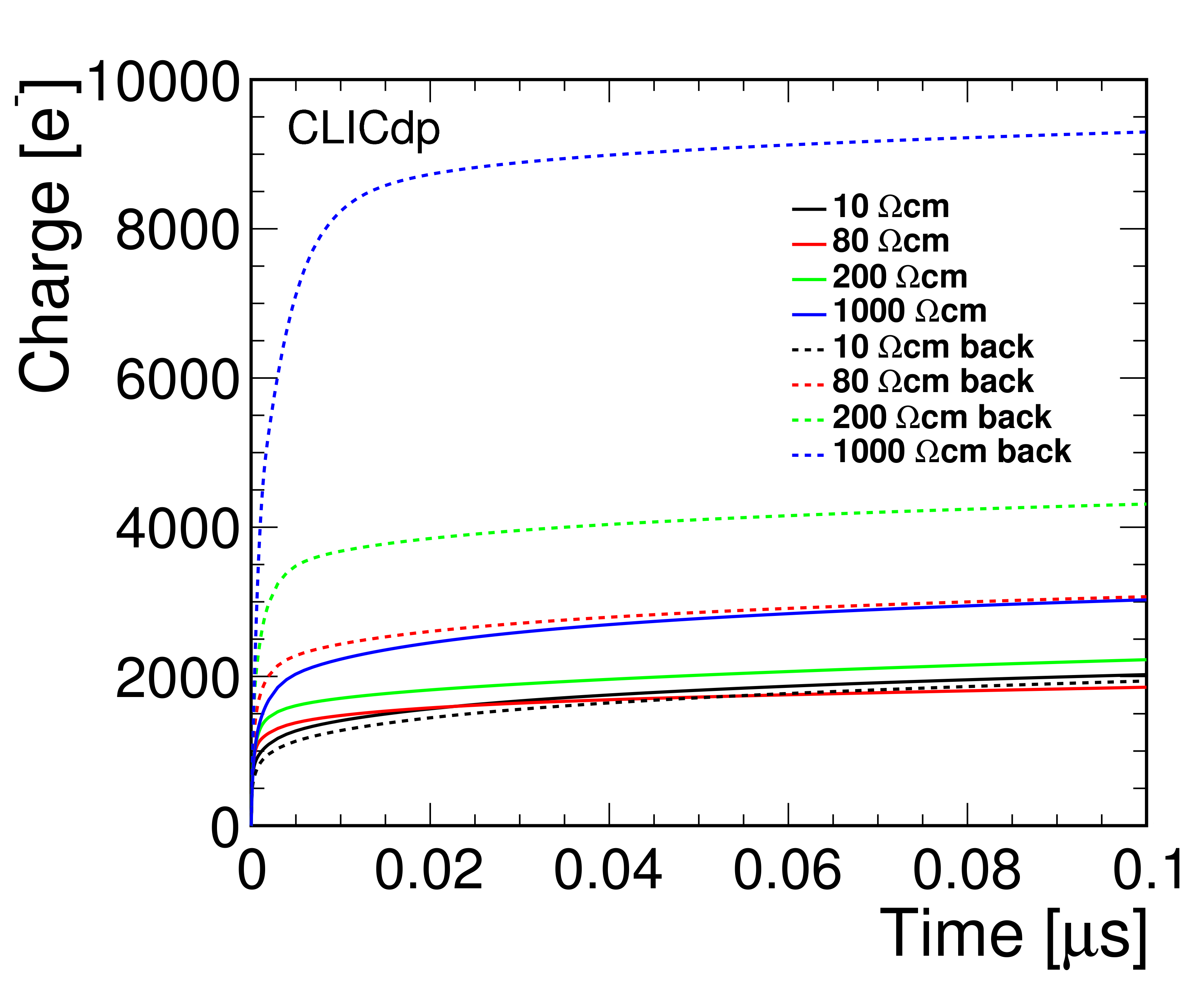}
	\end{tabular}	
	\caption{Comparison of the charge collection methods for (a) \textit{drift} and (b) \textit{diffusion}. (c) Charge collected for various bulk resistivities and the two biasing schemes. All plots are for a MIP passing perpendicular through the centre of the pixel cell at \SI{-60}{\volt}.}
	\label{fig: CC res comp}
\end{figure}

From the simulations, the breakdown voltage of high-resistivity substrates increase from around \SI{-88}{\volt} at \SI{10}{\ohm\cm} to around \SI{-100}{\volt} for higher values, while the leakage currents slightly increase due to the larger depletion region generating more carriers. This is shown in figure \ref{fig: IV res comp}, whereby comparing the two biasing schemes it can be seen that back biasing greatly improves the breakdown voltage. The charge collected gives an indication of the relative improvement between different resistivities and biasing schemes. Figure \ref{fig: CC res comp} shows the three charge collection modes: \textit{drift}, \textit{diffusion} and \textit{all}, as described in section \ref{sec:CCModes}, backing up the previous assumption that a larger resistivity will lead to larger collection from drift. The \textit{drift} curve for the \SI{1000}{\ohm\cm} model does not plateau but slightly increases with time because of the large generation rate for the \SI{1000}{\ohm\cm} model. It is interesting to note that the order is not the same for the \textit{diffusion} path, instead the \SI{10}{\ohm\cm} collects the largest amount after \SI{100}{\nano\s}. For the total charge, the \SI{1000}{\ohm\cm} model collects the largest amount of charge and differences between topside and backside biasing are again observed to increase with higher substrate resistivity (figure \ref{fig: CC res comp} (c)). One drawback of this backside biasing is however that extra processing has to be performed on the backside of the sensor, adding additional complexity and cost.

\section{Conclusions}

Measurements of HV-CMOS assemblies for the CLIC vertex detector have shown excellent tracking performance across the full angular acceptance, \SIrange{0}{80}{\degree}. Over this angle range, the single hit efficiency remains above 99\% and the single hit resolution is within \SIrange{5}{7}{\micro\m} after eta correction. For the device under test used in this study, the coupling between the HV-CMOS sensor and readout ASIC was not uniform across the matrix, owing to the limited amount of glue used during the assembly production. Despite this, sufficient coupling to see hits still existed between the sensor and readout chip in places where the glue was not present. TCAD simulations have been used to reproduce the performance of the sensor, showing general agreement with measurements of current-voltage, breakdown and charge collection characteristics. The simulations have also been used to optimise features of next generation sensor chips, and demonstrate that a move to higher resistivity substrates and backside biasing have benefits in depletion depth, breakdown and charge collection properties.

% add references

\printbibliography[title=References]

\end{document}